\PassOptionsToPackage{unicode}{hyperref}
\PassOptionsToPackage{hyphens}{url}
\documentclass[
]{article}
\usepackage{lmodern}
\usepackage{amssymb,amsmath}
\usepackage{ifxetex,ifluatex}
\ifnum 0\ifxetex 1\fi\ifluatex 1\fi=0 % if pdftex
  \usepackage[T1]{fontenc}
  \usepackage[utf8]{inputenc}
  \usepackage{textcomp} % provide euro and other symbols
\else % if luatex or xetex
  \usepackage{unicode-math}
  \defaultfontfeatures{Scale=MatchLowercase}
  \defaultfontfeatures[\rmfamily]{Ligatures=TeX,Scale=1}
\fi
\IfFileExists{upquote.sty}{\usepackage{upquote}}{}
\IfFileExists{microtype.sty}{% use microtype if available
  \usepackage[]{microtype}
  \UseMicrotypeSet[protrusion]{basicmath} % disable protrusion for tt fonts
}{}
\makeatletter
\@ifundefined{KOMAClassName}{% if non-KOMA class
  \IfFileExists{parskip.sty}{%
    \usepackage{parskip}
  }{% else
    \setlength{\parindent}{0pt}
    \setlength{\parskip}{6pt plus 2pt minus 1pt}}
}{% if KOMA class
  \KOMAoptions{parskip=half}}
\makeatother
\usepackage{xcolor}
\IfFileExists{xurl.sty}{\usepackage{xurl}}{} % add URL line breaks if available
\IfFileExists{bookmark.sty}{\usepackage{bookmark}}{\usepackage{hyperref}}
\hypersetup{
  pdftitle={Short-wave admittance correction for a time-domain cochlear transmission line model},
  hidelinks,
  pdfcreator={LaTeX via pandoc}}
\usepackage[margin=1in]{geometry}
\usepackage{longtable,booktabs}
\usepackage{etoolbox}
\makeatletter
\patchcmd\longtable{\par}{\if@noskipsec\mbox{}\fi\par}{}{}
\makeatother
\IfFileExists{footnotehyper.sty}{\usepackage{footnotehyper}}{\usepackage{footnote}}
\makesavenoteenv{longtable}
\usepackage{graphicx}
\makeatletter
\def\maxwidth{\ifdim\Gin@nat@width>\linewidth\linewidth\else\Gin@nat@width\fi}
\def\maxheight{\ifdim\Gin@nat@height>\textheight\textheight\else\Gin@nat@height\fi}
\makeatother
\setkeys{Gin}{width=\maxwidth,height=\maxheight,keepaspectratio}
\makeatletter
\def\fps@figure{htbp}
\makeatother
\providecommand{\tightlist}{%
  \setlength{\itemsep}{0pt}\setlength{\parskip}{0pt}}

\usepackage{authblk}
\usepackage{siunitx}
\usepackage[]{natbib}
\setcitestyle{aysep={},open={(},close={)}} 

\title{Short-wave admittance correction for a time-domain cochlear
transmission line model}

\author[1]{Fran\c cois Deloche \thanks{e-mail: francois.deloche@polytechnique.org}} % Write as First name Surname
\author[1]{Morgan Thienpont}
\author[1]{Sarah Verhulst}

\affil[1]{Hearing Technology @ WAVES, Department of Information Technology, Ghent University, Ghent, Belgium
}

\date{February, 2026}

\begin{document}
\maketitle

\begin{abstract}
Transmission line (TL) models implemented in the time domain can
efficiently simulate basilar-membrane (BM) displacement in response to
transient or non-stationary sounds. By design, a TL model is well-suited
for an one-dimensional (1-D) characterization of the traveling wave, but
the real configuration of the cochlea also introduces higher-dimensional
effects. Such effects include the focusing of the pressure around the BM
and transverse viscous damping, both of which are magnified in the
short-wave region. The two effects depend on the wavelength and are more
readily expressed in the frequency domain. In this paper, we introduce a
numerical correction for the BM admittance to account for 2-D effects in
the time domain using autoregressive filtering and regression
techniques. The correction was required for the implementation of a TL
model tailored to the gerbil cochlear physiology. The model, which
includes instantaneous nonlinearities in the form of variable damping,
initially presented insufficient compression with increasing sound
levels. This limitation was explained by the strong coupling between gain and frequency selectivity assumed in the 1-D nonlinear TL model, whereas cochlear frequency selectivity shows only a moderate dependence on sound level in small mammals. The correction factor was implemented in the gerbil model and made level-dependent using a feedback loop. The updated
model achieved some decoupling between frequency selectivity and gain,
providing 5 dB of additional gain and extending the range of sound
levels of the compressive regime by 10 dB. We discuss the relevance of
this work through two key features: the integration of both analytical
and regression methods for characterizing BM admittance, and the
combination of instantaneous and non-instantaneous nonlinearities.
\end{abstract}

\hypertarget{introduction}{%
\section{Introduction}\label{introduction}}

Cochlear macromechanics, simplified to a single degree of freedom such
as basilar membrane (BM) displacement, can be described by a
transmission line (TL) model capturing the essential features of the cochlear traveling wave. When incorporated into a pipeline that reproduces the various stages of
the auditory pathway (i.e., inner hair cells, auditory synapse and nerve,
brainstem nuclei) \citep{Verhulst2018}, TL models can compete with more
phenomenological models to generate evoked responses associated with the
auditory periphery \citep{Saremi2016, Vecchi2022}. Specific strengths of
the TL model include the simulation of otoacoustic emissions and the
reproduction of instantaneous nonlinearities. Simulations in the time
domain require a fine spatio-temporal sampling, but computations can be
accelerated through efficient numerical schemes \citep{Altoe2014} or,
most effectively, by training a deep neural network to reproduce the
mapping between sound input and BM output \citep{Baby2021}.

TL models effectively capture the traveling wave that propagates along
the BM in response to sounds, but their unidimensional (1-D) nature can
complicate the integration of effects that span multiple dimensions.
Some of these effects can be implicitly integrated into the 1-D
characterization of the traveling wave. For example, the property of
`scaling symmetry', which expresses the BM admittance as a function of
normalized frequency, has a formal connection with the tapering geometry
of the cochlea \citep{Shera1991}. This property ensures that the driving
pressure travels with a constant amplitude before reaching its peak,
akin to a planar wave \citep{Altoe2020}. Another aspect to consider is
the stability of the TL model, since the active cochlea has regions of
negative damping basal to each characteristic place that make cochlear
models prone to instability. In 2-D or 3-D models, stability can be
maintained by adding a viscous term dependent on the transverse gradient
of fluid velocity \citep{Sisto2021a, Sisto2023}. As the wave approaches
the peak region, this viscous term increases in magnitude relative to BM
velocity: the damping of the wave grows rapidly, naturally restraining
the region of negative damping. The case is more complicated for
time-domain TL models. An ingenuous workaround proposed by Zweig
\citep{Zweig1991} and further developed by Shera \citep{Shera2001}
involves adding a slow-acting, time-delayed stiffness to the BM to
restrain the region of negative damping to a region basal to the
characteristic place. This solution was integrated into a comprehensive
computational framework of the auditory periphery
\citep{Verhulst2012, verhulst2015, Verhulst2018}. The corresponding TL
model, referred to as the V-1D model, serves as the basis for the model
presented in this paper.

Although the BM admittance in the V-1D model includes a slow feedback
term, it remains based on a modified harmonic oscillator. As a result,
gain is tightly coupled to frequency selectivity, similar to a classical
second-order filter: a narrowly tuned BM admittance produces frequency
responses with a tall peak, while broader frequency responses are
associated with lower gain. This property is not a critical issue for
simulating human auditory responses because cochlear filters in humans
are believed to be significantly sharper than those of small mammals
\citep{Shera2002,Verschooten2018, Shera2019}. This sharper tuning aligns well with
the broad range of compression of the cochlear response. The problem is more apparent for models reproducing the auditory responses of common laboratory animals, which are of particular interest because they can be tested against a wider range of physiological data that require invasive techniques. For many of these animals (e.g., gerbils), at low sound levels where the cochlea
is in its most sensitive state, the peak of the BM velocity frequency
response is tall but relatively broad \citep{Fallah2021, He2022}.
Looking at multiple sound levels, the range of compression is large, but
the broadening of tuning is quite limited. This apparent discrepancy
posed a challenge during recent efforts to adapt the V-1D model to small
rodents, specifically gerbils and mice \citep{Thienpont2024}. In our initial attempt to
match cochlear responses in the gerbil, the model lacked approximately
10 dB of compression to achieve the desired variation in response gain.

In 2-D or 3-D models of the cochlea, part of the magnitude of the wave
peak can be attributed to the phenomenon of pressure focusing occurring
in the short-wave region. This effect is due to the decrease in the
characteristic height of the pressure field associated with the
shortening of the wavelength \citep{Reichenbach2014}. As the
characteristic height is reduced when approaching the wave peak, the pressure
becomes `focused' in a thin layer around the basilar membrane. This
translates into a positive gain factor on the pressure driving BM
velocity. This phenomenon is passive, but if cochlear compressive
nonlinearities are associated with an increase in wavelength, it has a
stronger effect in the most active state \citep{Sisto2021a}. In this
case, pressure focusing extends the range of the cochlear response gain
while having a limited impact on frequency selectivity.

In this paper, we use several numerical methods to simulate gain compensation from pressure focusing in a TL model tailored to the gerbil physiology. The
correction is based on the work by Sisto et al.~that integrates 2-D
dimensional fluid effects (viscous damping and pressure focusing) into a
frequency-domain TL model \citep{Sisto2021a}. We implement this
correction in the time domain using autoregressive filtering and
regression methods. Beyond the task-specific improvements presented
here, the new method may prove useful in various scenarios where
wavelength-dependent effects need to be accounted for in a TL model, a
challenge for time-domain implementations.

\hypertarget{methods}{%
\section{Methods}\label{sec:methods}}

In the Methods section, we use the following naming convention to refer
to the two transmission line (TL) models which we aim to combine in this work:

\begin{itemize}
\tightlist
\item
  The \textbf{V-1D model} refers to the time-domain TL implementation
  described in Verhulst et al.~\citep{Verhulst2018}, which builds upon Zweig and Shera's characterization of the
  BM admittance \citep{Zweig1991, Shera2001}.
\item
  The \textbf{S-2D model} designates the frequency-domain WKB
  approximation of the pressure wave proposed by Sisto et al
  \citep{Sisto2021a}. The model incorporates two 2-D hydrodynamic effects:
  pressure focusing and viscous damping, and serves as a reference for
  the short-wave admittance correction introduced in this paper.
\end{itemize}

Our contribution is the proposal of a third TL model, the
\textbf{\(V^{\star}\) model}, which is largely based on the
framework of the V-1D model, but incorporates some elements from the
S-2D model.

The Methods section is divided into two parts. The first three
subsections review key elements of the V-1D and S-2D models to establish a common framework for the \(V^{\star}\) model. The last subsection describes the implementation of the short-wave admittance correction.

\hypertarget{short-wave-gain-enhancement-from-pressure-focusing-s-2d-model}{%
\subsection{Short-wave gain enhancement from pressure focusing (S-2D
model)}\label{sec:S-2D}}

We begin with the presentation of the S-2D model, as it clarifies the connection between the TL equations and a 2-D physical model of the cochlear traveling wave. The equations integrate
pressure focusing and fluid viscous stress acting on the BM. More
details can be found in the original paper \citep{Sisto2021a}.

In the S-2D model, the upper and lower cochlear ducts are viewed as two
2-D rectangular boxes of height \(H\) separated by the BM. The
coordinates \(x\) and \(z\) correspond to the longitudinal and
transverse directions, respectively. We define \(p_d = p(z)-p(-z)\) as
the pressure difference between the scalae above the BM and the scala
tympani. \(U = \int_0^H u_x \, \mathrm{dz}\) is the volume velocity in
the upper scalae, and \(\overline{p_d}\) is the differential pressure
averaged along the transverse direction.

The ``TL-like'' equations characterizing the pressure wave are :

\begin{subequations}

\begin{align}
\frac{\partial \overline{p_d}}{\partial x}&= - Z_f \, U \label{eq:equa_motion} \\
\frac{\partial U}{\partial x}&= v_{BM}= Y_{BM} \, p_d(z=0^+) \,,  \label{eq:incomp0} 
\end{align}

\end{subequations}

where \(Z_f = \frac{2 j \omega \rho}{H}\), and \(Y_{BM}\) is the BM
admittance relating BM velocity to the pressure difference applied
across it.

The equations are almost reduced to a one-dimensional case, but in Eq.~\ref{eq:equa_motion}, the pressure \(\overline{p_d}\) represents an average along the
transverse direction, whereas in Eq.~\ref{eq:incomp0}, \(p_d\) is
evaluated at \(z=0^+\) which corresponds to the BM upper surface. By
introducing \(\alpha = p_d(z=0^+)/\overline{p_d}\), the `pressure
focusing' factor, the second equation can be rewritten as:

\begin{equation}\frac{\partial U}{\partial x}= Y_{BM} \, \alpha \, \overline{p_d} \ .\label{eq:incomp}\end{equation}

Assuming that \(Y_{BM} \, \alpha\) is a known function of the position
\(x\) and the pulsation \(\omega\), the 1-D formalism can be used. By
combining Equations \ref{eq:equa_motion} and \ref{eq:incomp}, we obtain
the wave equation:

\begin{equation}\frac{\partial^2 \overline{p_d}}{\partial^2 x} = - \kappa^2 \overline{p_d} \ ,\label{eq:wave_equa}\end{equation}

with \(\kappa^2 = Z_f Y_{BM} \alpha\).

By applying the WKB approximation to a 2-D box model, we obtain the following
formula for the pressure focusing factor:

\begin{equation}\alpha = \frac{\kappa H}{\tanh \kappa H} \ ,\label{eq:p_foc_factor}\end{equation}

leading to the dispersion relation \citep{Reichenbach2014}:
\begin{equation}\kappa \tanh(\kappa H) = 2 j \rho \omega \, Y_{BM} \, .\label{eq:disp_relation_0}\end{equation}

In Sisto et al., this approach is extended to include the effect of
fluid viscous stress acting on the BM. The viscous stress is
proportional to the transverse gradient of (transverse) fluid velocity
at \(z = 0\). In the classical potential flow description, the latter is
related to the second derivative of the pressure along \(z\), whereas
\(v_{BM}\) is related to the first derivative. The WKB approximation
gives \(\alpha/H\) for the ratio of the second to the first derivative.
The viscous stress acting on the BM, derived in detail in Sisto et al.,
is therefore:

\begin{equation}S = - 4 b \mu \, \frac{\alpha}{H} v_{BM}, \label{eq:visc_stress}\end{equation}

where \(\mu\) is the dynamic viscosity of the fluid and \(b=2.5\) is an
empirical factor to take into account the irrational part of the fluid
flow \citep{Sisto2021a} \footnote{Note that the empirical factor
\(b=2.5\) was not included in a more recent work \citep{Sisto2023}; here,
we kept this factor to maintain consistency with the 2021 paper.}. The BM
admittance is derived from a harmonic oscillator model, updated to
include the viscous stress:

\begin{equation}Y_{BM} = - \frac{j \omega}{\sigma_{BM}} \left[ -\omega^2+j \omega \, (\Gamma + \frac{4 \alpha b \mu}{\sigma_{BM} H}) + \omega_{bm}^2 \right]^{-1} \ .\label{eq:BM_admitt}\end{equation}

In this formula, \(\omega_{BM}\) is the characteristic frequency of the
BM, and \(\sigma_{BM}\) is the effective mass density of the BM.
\(\Gamma\) is a standard damping term, which can be further
decomposed into active and passive components:
\(\Gamma = \Gamma_a + \Gamma_p\) with
\(\Gamma_p = \omega_{BM}\) and
\(\Gamma_a = - G \omega_{BM}\). The variable \(G\) represents the
strength of the active process, which contributes to negative damping.

The updated BM admittance leads to the dispersion relation:

\begin{equation}\kappa \tanh(\kappa H)= \frac{2 \omega^2 \rho}{\sigma_{BM}}   \left[\omega_{bm}^2   -\omega^2+j \omega \, (\Gamma + \frac{4 \alpha b \mu}{\sigma_{BM} H}) \right]^{-1} \ .\label{eq:disp_relation}\end{equation}

This relation introduces an additional layer of complexity to the case
where \(Y_{BM}\) is independent of \(\kappa\), as the factor \(\alpha\), which depends on \(\kappa\)), appears on the right-hand side of the
equation. To compute \(\kappa\), the recursive procedure described in
Sisto et al.~is used. After convergence, the pressure focusing
factor \(\alpha\) is calculated using Eq.~\ref{eq:p_foc_factor}. For the
purpose of this paper, we also introduce the \(\beta\) factor, which
captures the variations of \(\alpha\) relative to a reference value:

\begin{equation}
\beta = \alpha/\alpha_0 \, ,
\label{eq:def_beta}
\end{equation} 
where \(\alpha_0\) is the pressure
focusing factor computed at a reference level. The S-2D parameters used
in our calculations were essentially the same as in Sisto et al. (reported in Table~\ref{tab:params}), but we adapted the value of \(H\) to
reflect the height of the gerbil cochlea at the mid-point between the base
and apex (\(CF=\SI{4}{\kilo \hertz}\)).

\hypertarget{time-domain-implementation-baseline-v-1d-model}{%
\subsection{Time-domain implementation: baseline (V-1D
model)}\label{sec:V-1D}}

Before presenting the strategy for integrating the pressure focusing
factor, we review some elements of the V-1D TL model that serves as the
basis for the time-domain implementation. The V-1D model consists of \(N=1000\)
cascaded sections representing \(N\) segments of the cochlear partition
from the base to the apex. Each section includes a series impedance
\(Z_{s_n}\) and a shunt (parallel) admittance \(Y_{p_n}\)
\citep{Zweig1991, Verhulst2012, Verhulst2018} :

\begin{equation}Z_{s_n}(j \omega)= M_{s_n} j \omega \ ,\label{eq:parallel_admittance}\end{equation}

\begin{equation}Y_{p_n} (j\omega)= \frac{j\omega}{M_{p_n}} \left[ \omega_{n}^2 - \omega ^2 + \delta_n j \omega \omega_n - \varrho_n e^{- j \psi_n  \omega/\omega_n} \omega_n^2  \right]^{-1} \ .\label{eq:shunt_admittance}\end{equation}

\(n\) indexes the TL sections from 0 (base) to \(N-1\) (apex). The
mass parameters scale as for a tapered guide with stiffness
proportional to the characteristic frequency
\citep{Zweig1991, Shera1991, Altoe2020} :
\(M_{s_n}= \frac{\omega_0}{\omega_n} M_{s_0}\), and
\(M_{p_n}= \frac{\omega_0}{\omega_n} M_{p_0}\) . The model is treated as
a system of coupled differential equations solved numerically using
Runge-Kutta (RK) 4(5), a finite-difference method.
\(\varrho_n e^{- j \psi_n \omega/\omega_n} \omega_n^2\) corresponds to a
slow feedback term, with stiffness \(\varrho_n\) and delay \(\psi_n\),
integrated into the RK4(5) scheme using a spline interpolation
\citep{Altoe2014}.

The V-1D model includes instantaneous nonlinearities that depend on the
magnitude of the BM velocity. Specifically, the triplet
\((\varrho_n, \delta_n, \psi_n)\) is updated over time to modify BM
damping while ensuring that the corresponding (double) pole trajectory
remains consistent with the near-invariance of BM velocity
zero-crossings \citep{Shera2001}. The formulas for determining the three
parameters as a function of BM velocity can be found in previous work
\citep{Verhulst2018}. The distance of the double pole from the imaginary
axis in the s-domain determines the damping of the local BM impulse
response. It is minimal at low sound levels and gradually increases to a
saturating value across a range of sound levels where the response is
compressive \citep{Verhulst2018}. This behavior mimics the
level-dependent nature of cochlear tuning and response gain. Eventually,
nonlinearities in the model are only determined by an initial state and
a parameter \(a\) controlling the compression of the response (in dB/dB)
as sound level increases. The initial state reflects the cochlea at low
sound levels, where the BM response is most sharply tuned and exhibits
the highest sensitivity \citep{Ruggero1998}. For animal models, the
compression parameter can be derived from BM-velocity growth functions
as the slope of the growth function represented on a logarithmic scale
\citep{Thienpont2024} .

The model version used in this study was calibrated to the gerbil physiology
(parameters in Table~\ref{tab:params}). It is the same as the gerbil model described in
\citet{Thienpont2024}, but the frequency
selectivity parameters were changed to reflect BM tuning rather than auditory nerve tuning. We made this choice to facilitate comparison of our model with BM vibration data. The poles in the V-1D model are set according to a power law relating the quality factor \(Q_{ERB}\) to center
frequency \citep{Verhulst2018}. In \citet{Thienpont2024}, this power law was matched to auditory
nerve data \citep{Muller1996}. In the
present study, we used the same power law except that \(Q_{ERB}\) was divided 
 by 1.9 to reflect BM tuning. This factor corresponds to the quality factor ratio between auditory-nerve and BM tuning reported in a recent study on cochlear frequency selectivity in the gerbil \citep{Charaziak2021}. 

\hypertarget{connections-between-the-two-tl-models}{%
\subsection{Connections between the two TL
models}\label{sec:compar_models}}

So far, we have introduced two existing TL models (V-1D and S-2D). At this stage of the
Methods section, it is useful to highlight their similiarities and differences, 
and to emphasize that each serves a distinct purpose in the rest of the
paper. Both models aim to characterize the cochlear traveling wave and
its nonlinear behavior with sound level. They are based on coupled equations of
pressure and velocity, which have been presented in the form of partial
differential equations for the S-2D model (Eqs.~\ref{eq:equa_motion} to
\ref{eq:incomp}), and in the form of a system of impedances for the V-1D
model (Eqs.~\ref{eq:parallel_admittance} and \ref{eq:shunt_admittance}).
One can therefore note many formal similarities between the two model,
although different notations were used to avoid confusion between
their respective parameters.

The two models, however, differ in several ways. The S-2D model is based
on physical principles. This model is easier to solve numerically in the
frequency domain, but the BM impedance does not have a closed-form
expression, even in the frequency domain. By contrast, the V-1D model
was specifically designed to provide a relatively simple expression for
the BM admittance and to reproduce time-domain characteristics such as
the invariance of zero-crossings \citep{Shera2001}. It is therefore more natural to consider the V-1D model as the basis of the time-domain
implementation. The slow-delay feedback term in the admittance of V-1D
model is not connected to any known physiological mechanism in the
cochlea, but provides the necessary frequency modulation to alternate
between negative and positive resistance regions in the frequency domain, ensuring numerical stability.

The two models also differ in their behavior when simulating
traveling-wave responses, particularly in how the region of the wave
peak is represented. In the V-1D model, the peak occurs at the transition between the stiffness- and mass-dominated regions, while in the
S-2D model, the peak occurs at a more basal place (i.e., in the stiffness-dominated region). This
difference will be stressed further in the Discussion section. Here, we simply note that it leads to different intepretations of the local BM resonant frequency in each model. In the V-1D model,  the
peak of the traveling wave in response to a tone coincides with the region of characteristic pulsation \(\omega_n= 2\pi f\).  In 
other words, at low sound levels the characteristic frequency \(CF\) and the best frequency
\(BF\) are identical \citep{Shera2001}. In the S-2D
model, the wave is slowed down and peaks before it reaches its
the characteristic place defined by the ratio of mass and stiffness
\citep{Sisto2021a, Sisto2024b}, so that we have \(BF < CF\) at the location of the peak. In practice, this discrepancy is
resolved by applying an empirical affine relation to \(\omega_{BM}\) so
that the best frequencies of each model coincide. Based on WKB
simulations of the S-2D model, we determined that
\(\omega_{BM} = 1.2 \, \omega_n + \SI{1500}{Hz}\), where \(\omega_{BM}\)
is the characteristic pulsation of the S-2D model, and the best
frequency is equal to \(\omega_n/(2 \pi)\). This relationship is used to
assign a value of \(\omega_{BM}\) to each TL section for the calculation
of the pressure focusing gain. The place-frequency mapping is set as in
the V-1D model by a Greenwood function:

\begin{equation}
\omega_n= 2\pi A_1 (10^{-A_2 x}-B) \ ,
\label{eq:greenwood_mapping}
\end{equation} 
where \(x\) is the distance from
base (in meter). The S-2D model is only used for computing the
pressure focusing factor, which is a local property, so that we do not
need to set a tonotopic mapping for S-2D to implement the methods described in the following subsections.

\hypertarget{implementation-of-the-admittance-correction-vstar-model}{%
\subsection{\texorpdfstring{Implementation of the admittance correction
(\(V^\star\)
model)}{Implementation of the admittance correction (V\^{}\textbackslash star model)}}\label{sec:implementation}}

\begin{figure}
\hypertarget{fig:setup_phase}{%
\centering
\includegraphics{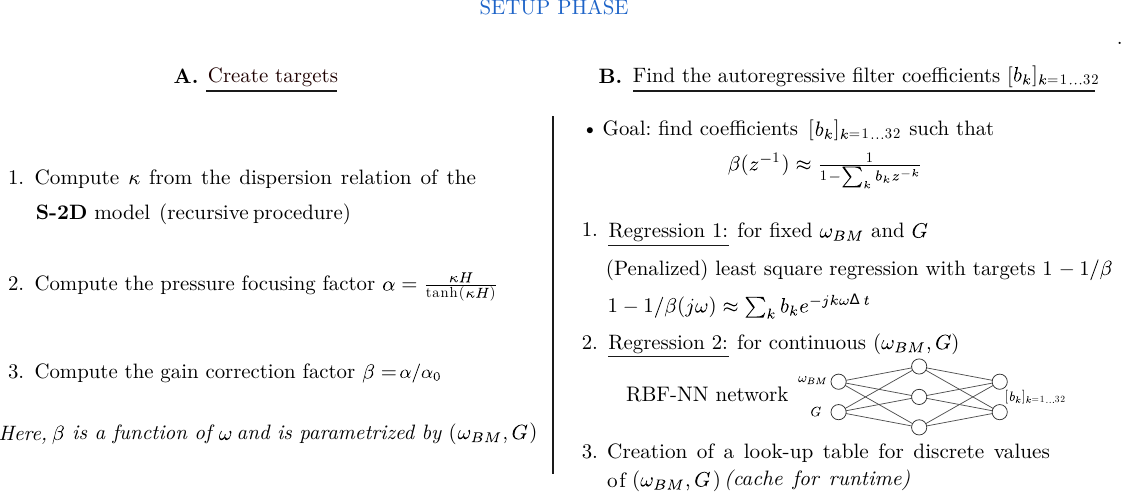}
\caption{Implementation strategy for the setup of the admittance
correction.}\label{fig:setup_phase}
}
\end{figure}

\begin{figure}
\hypertarget{fig:runtime_phase}{%
\centering
\includegraphics{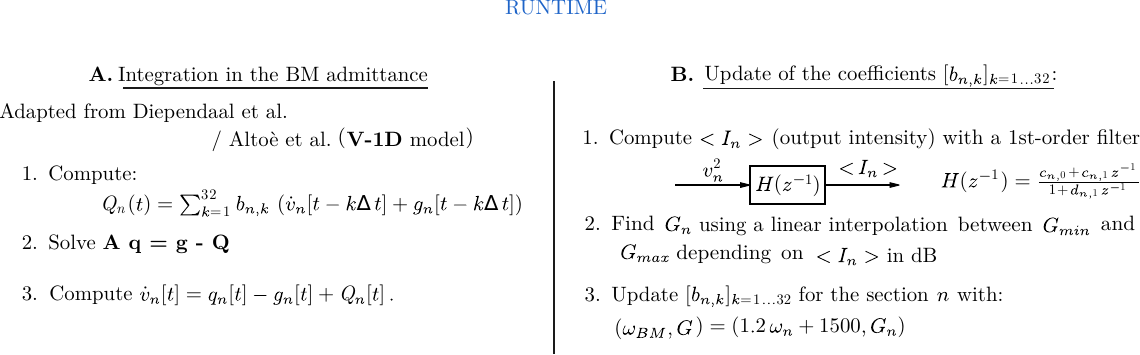}
\caption{Implementation strategy at runtime (when running simulations).}\label{fig:runtime_phase}
}
\end{figure}

Figures \ref{fig:setup_phase} and \ref{fig:runtime_phase} list the major
steps of the implementation, divided into setup phase and runtime.

The general strategy is as follows:

\begin{enumerate}
\def\labelenumi{\arabic{enumi}.}
\tightlist
\item
  Create a function \(\beta=\beta(\omega_{BM}, G)\) that returns the
  correction factor to apply to the BM admittance, computed from the
  S-2D model. The inputs of this function are the BM characteristic
  frequency and the strength of the `active process'. The output is a
  vector defined in the frequency domain.
\item
  Determine autoregressive coefficients to approximate
  \(\beta\) as the frequency response of an all-pole filter. The filter
  coefficients are denoted by \([b_{k}]_{k=1 \cdots 32}\), and also depend on \(\omega_{BM}\) and \(G\). This step includes two regression
  methods that are run consecutively.
\item
  Update the time-domain equations to apply the recursive filters
  \(\beta\), defined by the coefficients \([b_{n, k}]_{k=1 \cdots 32}\),
  on top of the V-1D BM admittance.
\item
  During simulation: update the parameters
  \([b_{n, k}]_{k=1 \cdots 32}\) over time to reflect changes in the
  active process, whose strength varies with the response intensity.
\end{enumerate}

Steps 1 and 2 (setup phase) only needs to be run once to find the
regression parameters defining the recursive filters. Steps 3 and 4 are
executed when running the simulations. The following subsections describe each step in
detail.

\hypertarget{estimation-of-the-autoregressive-filter-coefficients}{%
\subsubsection{Estimation of the autoregressive filter
coefficients}\label{sec:autoregressive-filter}}

Our goal is to implement the gain missing in the V-1D model by
introducing the pressure focusing factor \(\alpha\) computed from the
S-2D model (Eq.~\ref{eq:p_foc_factor}). However, since \(\alpha\) can
take large values that are difficult to incorporate directly in a
time-domain implementation, we instead use the normalized \(\beta\)
factor, introduced previously (Eq.~\ref{eq:def_beta}: \(\beta= \alpha/\alpha_0 \)).
\(\alpha_0\) is the pressure focusing factor computed using the
reference value \(G=G_{REF}\) for the strength of the active process. \(G_{REF}\) is chosen as the midpoint of
the range of values for \(G\) (\([G_{\min}, G_{\max}]\)), to keep \(\beta\) as close as
possible to 1. For clarity, we omit the full dependencies of
\(\beta\), which should be written as \(\beta(\omega_{BM}, G, \omega)\).
%Throughout the rest of the paper, \(\beta\) is treated as a
%frequency-dependent vector, with its input variables (most often
%implicit), defined by the pair \((\omega_{BM}, G)\). 
We call \(\beta\) a gain enhancement or correction factor, although in practice it can have a magnitude lower than 1. Note also that \(\beta\) has complex outputs, i.e., \(\beta\) is defined by its magnitude but also phase.

% \begin{longtable}[]{@{}ll@{}}
% \toprule
% \textbf{S-2D model} (calculation of \(\alpha\) and \(\beta\)) &
% {[}{]}\tabularnewline
% \midrule
% \endhead
% \textbf{H} & \(\mathbf{\SI{320}{\micro \meter}}\)\tabularnewline
% \(\sigma_{BM}\) & \(\SI{0.06}{\kilo\gram\per\meter^2}\)\tabularnewline
% \(\rho\) & \(\SI{1.0e3}{\kilo\gram\per\meter^3}\)\tabularnewline
% \(\mu = 10 \mu_w\) & \(\SI{7.0e3}{\milli\pascal\second}\)\tabularnewline
% \(\mathbf{[G_{\min}, G_{\max}]}\) & \textbf{{[}0, 1.3{]}}\tabularnewline
% \(\mathbf{G_{REF}}\) & \textbf{0.7}\tabularnewline
% \textbf{V-1D / \(\mathbf{V^\star}\) model} (time-domain implementation)
% &\tabularnewline
% \(f_s\) (sampling frequency) & \(\SI{200}{\kilo \hertz}\)\tabularnewline
% \(N\) (number sections) & 1000\tabularnewline
% H & \(\SI{320}{\micro \meter}\)\tabularnewline
% Greenwood \(A_1\) & \(\SI{50216}{\hertz}\)\tabularnewline
% Greenwood \(A_2\) & \(\SI{181.034}{\per\meter}\)\tabularnewline
% Greenwood B & \(\SI{140}{\hertz}\)\tabularnewline
% \(Q_{ERB}\) (\(f_{0}=\) 1 kHz) &
% \(1.45 \cdot (f/f_{0})^{0.58}\)\tabularnewline
% \(a\) (compression slope) & 0.45 dB/dB\tabularnewline
% \textbf{Regressions} &\tabularnewline
% \(\lambda_1\) (first LSE regression penalty) & 1\tabularnewline
% \(\lambda_2\) (second LSE regression penalty) & 0.3\tabularnewline
% RBF-NN parameters & in text.\tabularnewline
% \bottomrule
% \end{longtable}

\begin{table}[h]
    \centering
    \caption{Model and regression parameters. \textbf{Top:} Parameters
for the S-2D model are the same as in \citep{Sisto2021a} (WKB
simulations with \(\mu=10\mu_w\)), except for the parameters in bold.
\textbf{Middle:} Main parameters for the V-1D and \(V^\star\) models.
For more details on the model parametrization, refer to \citet{Thienpont2024}. 
\vspace{1em}}
    \label{tab:params}
    \begin{tabular}{ll}
        \toprule
        \textbf{S-2D model} (calculation of \(\alpha\) and \(\beta\)) & \\
        \midrule
        \textbf{H} & \(\mathbf{\SI{320}{\micro \meter}}\) \\
        \(\sigma_{BM}\) & \(\SI{0.06}{\kilo\gram\per\meter^2}\) \\
        \(\rho\) & \(\SI{1.0e3}{\kilo\gram\per\meter^3}\) \\
        \(\mu = 10 \mu_w\) & \(\SI{7.0}{\milli\pascal\second}\) \\
        \(\mathbf{[G_{\min}, G_{\max}]}\) & \textbf{{[}0, 1.3{]}} \\
        \(\mathbf{G_{REF}}\) & \textbf{0.7} \\
        \midrule
        \textbf{V-1D / \(\mathbf{V^\star}\) model} (time-domain implementation) & \\
        \midrule
        \(f_s\) (sampling frequency) & \(\SI{200}{\kilo \hertz}\) \\
        \(N\) (number of sections) & 1000 \\
        H & \(\SI{320}{\micro \meter}\) \\
                L (cochlear length) & \SI{12.1}{\milli \meter} \\
        Greenwood \(A_1\) & \(\SI{50216}{\hertz}\) \\
        Greenwood \(A_2\) & \(\SI{181.034}{\per\meter}\) \\
        Greenwood B & \(\SI{140}{\hertz}\) \\
        \(Q_{ERB}\) (\(f_{0}=\) 1 kHz) & \(1.45 \cdot (f/f_{0})^{0.58}\) \\
        \(a\) (compression slope) & 0.45 dB/dB \\
        \midrule
        \textbf{Regressions} & \\
        \midrule
        \(\lambda_1\) (first LSE regression penalty) & 1 \\
        \(\lambda_2\) (second LSE regression penalty) & 0.3 \\
        RBF-NN parameters & in text. \\
        \bottomrule
    \end{tabular}
\end{table}

For a given pair \((\omega_{BM}, G)\), we approximate \(\beta\) as the
frequency response of an all-pole filter. Specifically, we want to find
coefficients \([b_k]_{k=1 \cdots 32}\) such that:

\begin{equation}\beta(z^{-1}) \approx \frac{1}{1-\sum_{k=1}^{32} b_k z^{-k}} .\label{eq:approx1}\end{equation}

We used 32 coefficients as a trade-off between accuracy and
computational cost, but the method is general and can work with
another number of coefficients. In the frequency domain, we can also
write

\begin{equation}1-1/\beta(j \omega) \approx  \sum_{k=1}^{32} b_k e^{-j k \omega \Delta t} \ .\label{eq:approx_lin_reg}\end{equation}

We use this formula to find the coefficients \(b_k\). We discretize the equality over regularly spaced frequencies, representing each side as a vector in the frequency domain. We then determine the coefficients \(b_k\) that minimize the least-squares error (LSE) between these two vectors. We limit the computation
of the LSE to frequencies up to \(\omega_{cut} = 1.3 \, \omega_{BM}\) ,
as frequencies above are filtered out by the transmission line. However,
we add two penalty terms that ensure that: a) the coefficients
\([b_k]_{k=1 \cdots 32}\) remain small; b) The function
\(\hat{\beta}(j \omega) = \left[1 - \sum_{k=1}^{32} b_k e^{-j k \omega \Delta t} \right]^{-1}\)
has a small or negative imaginary part relative to its real part for
frequencies above \(\omega_{cut}\) .

The first condition is enforced by applying a penalty on the sum of
squares of \([b_k]_{k=1 \cdots 32}\), as in ridge regression. The second
condition is controlled by penalizing
\(\max\left(0, \mathrm{Im}(\hat{\beta})/\mathrm{Re}(\hat{\beta})\right)\)
above \(\omega_{cut}\), with the \(\max\) operator applied element-wise.
This penalty prevents situations where a monochromatic solution of the
TL model, locally proportional to \(\exp(- j \kappa x)\), becomes
unstable at high frequencies due to a positive imaginary part in
\(\kappa\).

Appendix \ref{appendix:cost-function} provides more details on the LSE regression used to
determine the \([b_k]\) coefficients, including the full form of the
cost function. We also describe a method that we used to enforce null gain of the correction factor at low frequencies, in Appendix~\ref{appendix:cost-function-withgain}. This precaution was taken to ensure that a low-frequency wave travels up to its peak as a planar wave (i.e., with a constant characteristic admittance)]\cite{Altoe2020}, since the opposite can lead to spurious behaviors \citep{Shera1991}.

The regression method presented above works for a given pair
\((\omega_{BM}, G)\). To achieve a continuous interpolation of the
\([b_k]\) coefficients when \((\omega_{BM}, G)\) is allowed to vary, we
applied a second regression method using a radial basis function neural
network (RBF-NN). The input of the network is a tuple
\((\omega_{BM}, G)\) and the output is the set of estimated coefficients
\([b_k]_{k=1 \cdots 32}\) that approximate \(\beta\) according to 
Eq.~\ref{eq:approx1}. The following sentences provide more details
on the RBF-NN and how it was trained. The inputs were first normalized
to have input values between 0 and 1. There were 360 RBF centers
(\(18 \times 20\)), each associated with a Gaussian kernel of standard
deviation 0.04. The kernel was made `local linear' in the first
dimension, meaning it was multiplied by an affine function with variable
\(\omega_{BM}\). The RBF-NN was trained using stochastic gradient
descent on randomly generated pairs \((\omega_{BM}, G)\). Two strategies
were considered for defining the cost function. In the first, targets
\([b_k]\) are generated using the first method (penalized linear
regression), and the Euclidean distance between the targets and the
RBF-NN outputs is minimized. The second strategy directly minimizes the
LSE cost function (on Eq.~\ref{eq:approx_lin_reg}), bypassing the
generation of \([b_k]\) coefficients from the first method. We trained
the RBF-NN with 1,000 steps using the first strategy (learning rate:
0.1, momentum: 0.8) and 1,000 subsequent steps to fine-tune the
parameters using the second strategy (learning rate: 1e-3, momentum:
0.9).

When simulating the traveling wave in response to a stimulus, generating
the \(\beta\) coefficients on the fly using the RBF-NN can be
computationally intensive. To reduce the computation time, we allowed
the beta coefficients to be pre-computed and stored in a lookup table of
size \(1000 \times 30\), where \(N=1000\) is the number of channels and
30 is the number of discrete steps covering the range of \(G\) values.
The lookup table needs to be created only once, after which it can be reused for all simulations.

\hypertarget{integration-of-the-new-filters-in-the-v-1d-model}{%
\subsubsection{Integration of the new filters in the V-1D
model}\label{sec:time-domain-equations}}

\emph{Note: in this subsection, we add the subscript \(n\) to the
coefficients \([b_{n, k}]\) to emphasize the dependence of the filter
coefficients on the TL section indexed by \(n\)}.

The shunt admittance in the TL equations (Eq.~
\ref{eq:shunt_admittance}) now includes the gain correction factor
\(\beta(j \omega)\) but the time-domain equations need to be updated
accordingly. We introduce, as in \citet{Altoe2014},
the variables:

\[g_n(t) = \omega_{n}^2 y_n(t) + \delta_n \omega_n v_n(t) - \varrho_n \omega_n^2 y_n(t-\tau_n), \]

\[q_n(t)= \frac{1}{L_{BM} M_{p_n}} p_n(t), \]

where \(\tau_n = \psi_n/\omega_n\) and \(L_{BM}\) is the BM width. The
law of motion including the pressure focusing factor gives:

\begin{equation}
\dot q_n= (\dot v_n + g_n)\cdot (1 - \sum_{k=1}^{32} b_{n, k} z^{-k}) \, .
\end{equation}

It is possible to separate current and past samples to get:

\begin{equation}
\begin{aligned}
\dot v_n[t] &= q_n[t] - g_n[t] + \mathit{Q}_n[t], \text{where}\\
\mathit{Q}_n[t] &= \sum_{k=1}^{32} b_{n,k} \left( \dot{v}_n[t - k \Delta t] + g_n[t - k \Delta t] \right) \, .
\end{aligned}
\label{eq:main_time_equation}
\end{equation}

Equation~\ref{eq:main_time_equation} is very similar to the update equation in 
\citet{Diependaal1987} and \citet{Altoe2014}, but with a dependence on past
samples introduced through the new variable \(\mathit{Q}_n(t)\). The
second set of equations is derived from the relation between
\(\Delta p\) and \(\dot{v_n}(t)\) \citep{Diependaal1987}, which can be
written as:

\begin{equation}\mathbf{A q = g - Q}\label{eq:tridiag}\end{equation}

where \(\mathbf{A}\) is a tridiagonal matrix (see for example Eq.~2 in
\citet{Verhulst2012}), and \(\mathbf{q}\), \(\mathbf{g}\), and
\(\mathbf{Q}\) are the vector forms of \([q_n]\), \([g_n]\), and
\([\mathit{Q}_n]\) .

The simulation of BM motion follows the same approach as in~\citet{Diependaal1987}, by solving Eq.~\ref{eq:tridiag} then
Eq.~\ref{eq:main_time_equation} at each step. To improve accuracy and
efficiency, the V-1D model employs a RK4(5) adaptive numerical scheme
\citep{Altoe2014}, which requires interpolating past values using cubic
splines to account for the long-delay feedback term. We adopt the same
strategy to provide an estimation of \(Q_n\) at each inner step. A
complication is that \(\mathit{Q}_n\) contains a term depending on
\(\dot{v}_n[t - \Delta t] + g_n[t - \Delta t]\) (for \(k=1\)), which
cannot be interpolated using cubic splines because the value of
\(\dot{v}_n[t + \Delta t]\) is not yet known at timestep \(t\). To
address this, we split \(\mathit{Q_n}\) into two parts: the first term,
\(b_{n, 1} \left( \dot{v}_n[t - \Delta t] + g_n[t - \Delta t] \right)\),
is estimated using Lagrange's quadratic interpolation with the points at
\(t - 2 \Delta t\), \(t - \Delta t\), and \(t\). The second term,
\(\sum_{k=2}^{32} b_{n, k} \left( \dot{v}_n[t - k \Delta t] + g_n[t - k \Delta t] \right)\),
is interpolated with Catmull-Rom cubic splines using the same points
and \(t + \Delta t\).

\hypertarget{update-of-the-filter-coefficients-during-simulation}{%
\subsubsection{Update of the filter coefficients during
simulation}\label{sec:update-filter-coefficients}}

The correction gain factor \(\beta\) depends on the strength of the
active process (variable \(G\) in the S-2D model) which varies over time
as a function of response intensity. For each cochlear section, we
compute a short-term average of intensity using a first-order recursive
filter:

\begin{equation}
\langle I_n \rangle[t] = c_{n,0} \, v_n^2[t] + c_{n, 1} \, v_n^2[t - \Delta t] - d_{n, 1} \, \langle I_n \rangle[t - \Delta t] \ .
\end{equation}
The coefficients \([c_n], [d_n]\) were chosen to implement a first-order low-pass
Butterworth filter with a cut-off frequency of \(f_n/2 = \omega_n/(4\pi)\). This cut-off ensures that the envelopes \(\langle I_n \rangle(t)\) are smooth yet still react rapidly to changes in the signal intensity.

The \(V^\star\) model retains the nonlinearities of the V-1D model that
depend on the instantaneous velocity according to three different
regimes: a linear regime up to a first transition point \(v_{knee1}\), a
compressive regime between \(v_{knee1}\) and a second transition point \(v_{knee2}\), and a
second linear regime above \(v_{knee2}\). The trajectory of the BM admittance (double)
pole value in the compressive regime is set by a constant compression
slope, denoted as \(a\), using a hyperbolic interpolation
\citep{Verhulst2018}.

The \(V^\star\) and V-1D model differ slightly in how the transition
points are determined. The knee points are set using simulations of tone responses at the two linear regimes defined above. The poles are called the starting and saturating poles for the regime at low and high sound
levels, respectively. For the gerbil model, we plotted the growth functions of \(v_{BM}\) at the CF=20 kHz place. Because the two regimes are linear, the growth functions are represented by straight lines when \(v_{BM}\) is plotted against the tone level (both values in dB). The growth function obtained with the starting poles is used
to determine \(v_{knee1}\), corresponding to \(I_{knee1\_dB}\), the
sound level marking the onset of compression. The second transition
point \(v_{knee2}\) is determined using the simulations with the
saturating pole, by finding the intersection of the growth function with
the line with equation \(v_{dB}=v_{knee1\_dB}+a (I_{dB}-I_{knee1\_dB})\) where \(a\) is the compression slope. To
calibrate the \(V^\star\) model, we follow the same procedure, but the
pressure focusing factor is set to its maximal value (\(G=G_{\max}\)) in the case of the starting pole, and to its minimal
value (\(G=G_{min}\)) in the case of the saturating pole. The
calibration of compressive nonlinearities is illustrated at the end
of the Results section.

Once the transition points are set, the pressure focusing factor,
depending on \(G\), can be updated dynamically. Within the compressive
regime, \(G\) is determined by a linear interpolation between \(G_{\min}\) and  \(G_{\max}\) with velocity
values expressed in dB:

\[G(t) = G_{\max} + \frac{\langle I_n \rangle_{dB}(t) -6\, dB - v_{knee1\_dB}}{ v_{knee2\_dB} - v_{knee1\_dB}} (G_{\min} - G_{\max}) \ .\]

The -6 dB factor is an empirical factor to align \(\langle I_n \rangle\)
with the RMS values \(v_{knee1, 2}\). In the gerbil model simulations,
the coefficients \([b_k]\) were updated
according to this formula at every 6 timesteps (i.e., every 0.03 ms with
\(f_s=\SI{200}{\kilo \hertz}\)).

\hypertarget{results}{%
\section{Results}\label{sec:results}}

\hypertarget{compressive-response-in-the-v-1d-model}{%
\subsection{Compressive response in the V-1D
model}\label{sec:compressive-response-in-the-v-1d-model}}

\begin{figure}
\hypertarget{fig:initial_model}{%
\centering
\includegraphics{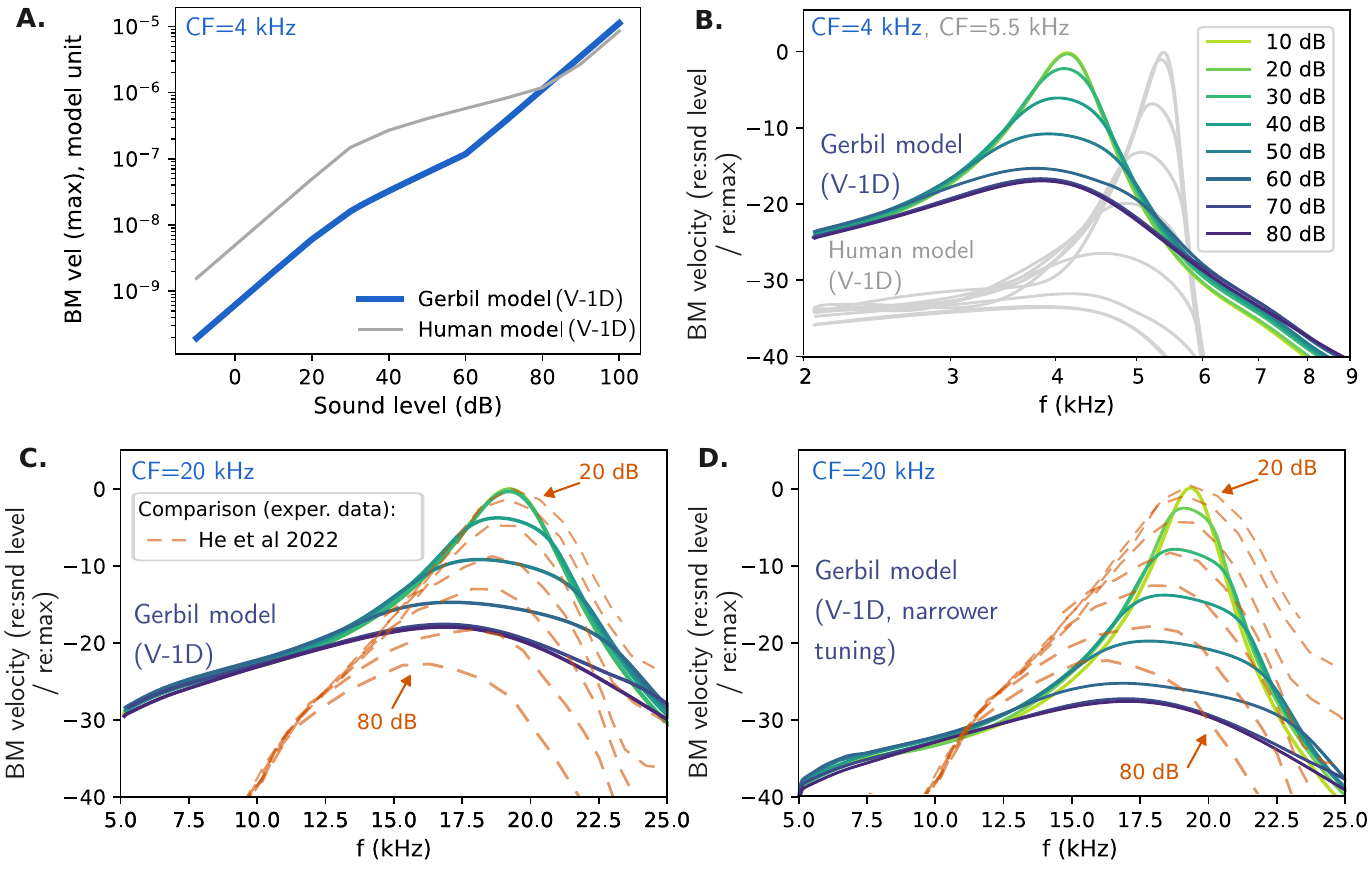}
\caption{\textbf{Characteristic responses of the initial V-1D model.}
\textbf{A.} BM velocity growth functions with respect to stimulus level
at the place with CF=4 kHz (the stimulus is a CF tone).
\textbf{B to D.} Frequency responses (magnitudes) of the V-1D model estimated by the
BM velocity responses to a slow linear chirp presented at
different sound levels. \textbf{B} Frequency response at the CF= 4 kHz place for the gerbil model
(in color). Responses for the human model but at CF=5.5 kHz are also 
shown (in gray). \textbf{C.} Frequency response at the CF=20 kHz place. The color code for sound level is the same as in B. Vibration data from OCT measurements
in gerbils \citep{He2022} are overlayed with dashed lines. For the experimental data, sound levels
are in dB SPL and velocity is relative to motion at the stapes. \textbf{D.} Same as C., but with a model version
with sharper tuning.
Figures B. to D.: `re:snd level / re:max' indicates that the frequency
responses were normalized to the the sound level and to the maximum peak magnitude of each curve set.}\label{fig:initial_model}
}
\end{figure}

We begin by presenting the issue of insufficient compression in the
initial V-1D model adapted to the gerbil physiology. Figure
\ref{fig:initial_model}A shows the BM velocity growth for a tone stimulus. The simulations correctly feature a compressive regime flanked
by two linear regimes at low and high stimulus levels. However, the
range of sound levels over which compression occurs is limited to less
than 30 dB. For the model version tailored to human hearing
\citep{Verhulst2018}, the compressive regime spans 40 dB and presents a stronger compression slope. The origin of these distinct
behaviors becomes clearer with the frequency responses shown in Fig.~
\ref{fig:initial_model}. These frequency responses were obtained by
simulating BM responses to a slow linear chirp (2--10 kHz or 5--30 kHz), and analyzing the response with a sliding Gaussian
window. In Fig.~ \ref{fig:initial_model}B, responses for the gerbil and
human models are shown next to each other with comparable CFs (gerbil: 4
kHz, human: 5.5 kHz). It shows that the human model features a much
sharper frequency tuning along with a broader range of response gain across
sound levels. In the V-1D model, these two properties are coupled so that reducing the frequency selectivity results in less compression available.

Figure \ref{fig:initial_model}C confirms these observations at a higher CF and shows a comparison with published BM vibration data. In Fig.~\ref{fig:initial_model}C and \ref{fig:initial_model}D
 we reproduced the frequency response measured in the basal turn of a live gerbil's cochlea using
optical coherence tomography (OCT) \citep{He2022}. The best frequency for the experimental data and the simulations is 19.5 kHz; for simplicity, we consider that CF=20 kHz throughout the text. The overlay with V-1D
model simulations in Fig.~\ref{fig:initial_model}C confirms similar
frequency tuning, but the compression at the best frequency is weaker in the model by
approximately 10 dB. The full range of compression can be restored in anW
alternative model version with sharper tuning
(Fig.~\ref{fig:initial_model}D), but this comes at the cost of unrealistically high frequency selectivity at low sound levels. The
trade-off between gain and frequency selectivity can shift the issue
from insufficient compression to inadequate tuning, but the V-1D model
cannot resolve both aspects simultaneously.

\hypertarget{short-wave-gain-enhancement-in-the-vstar-model}{%
\subsection{\texorpdfstring{Short-wave gain compensation in the
\(V^\star\)
model}{Short-wave gain enhancement in the V\^{}\textbackslash star model}}\label{sec:short-wave-gain-enhancement-in-the-vstar-model}}

\begin{figure}
\hypertarget{fig:regression_results}{%
\centering
\includegraphics{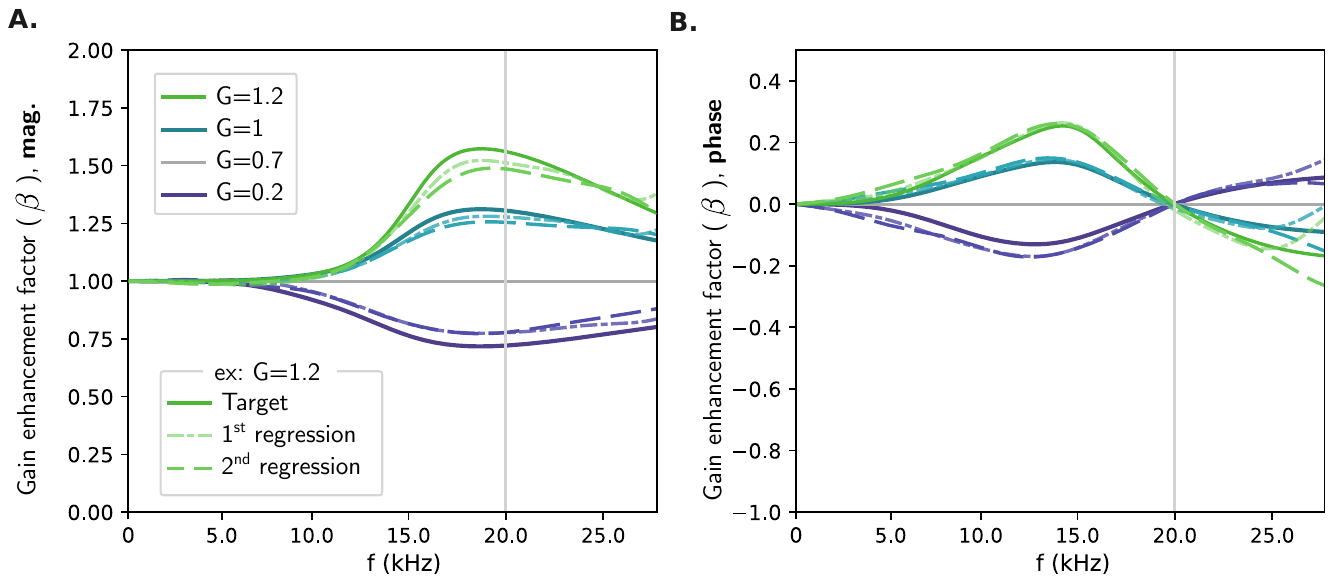}
\caption{Gain enhancement factor \(\beta\) (\textbf{A.} gain and
\textbf{B.} phase) for four different values of \(G\) (strength of the
active process). The target was calculated using the recursive procedure
described in Sisto et al.~for the S-2D model and the relation
\(\beta = \alpha/\alpha_0\) where \(\alpha_0\) is the pressure focusing
factor for \(G_{REF}=0.7\) (reference value). The dashed curves
correspond to the regressions used to approximate \(\beta\) as a
recursive all-pole filter. Dash-dotted: first regression (ridge
regression). Dashed: second regression (RBF neural network). In the two
panels, the vertical line corresponds to the BM characteristic frequency
\(\omega_{BM}/(2\pi)\).}\label{fig:regression_results}
}
\end{figure}

The \(V^\star\) model introduces partial gain compensation through the addition of a recursive all-pole filter (denoted by \(\beta\)), acting on top of the BM admittance. The magnitude and phase of
\(\beta\) for different strengths of the active process are shown in
Fig.~\ref{fig:regression_results}. As expected, \(\beta\) has virtually no effect (gain of 0 dB) for low frequencies corresponding to the long-wave region, but exhibits stronger variations for short-wave frequencies (e.g., close to CF). The model only implements a version of the filter obtained through two successive regressions. The result of the second regression (RBF-NN) can deviate by 25\% in
magnitude from the initial target (S-2D model). The regression
underestimates \(\beta\) if it represents a gain enhancement from the
reference configuration, while it overestimates \(\beta\) in case of a
gain reduction (i.e., \(\beta<1\)). This indicates that the method is
limited to relatively small corrections of the BM admittance.
Nonetheless, the overall trend of \(\beta\) across frequencies and \(G\)
values is well captured. The stability of the resulting filters was
inspected by calculating numerically their poles and representing them
in the z-domain. No case of instability was observed. The stability of the filters is promoted by the small filter coefficients that result from the ridge-like regression.

\begin{figure}
\hypertarget{fig:v_star_response}{%
\centering
\includegraphics[width=3.5in]{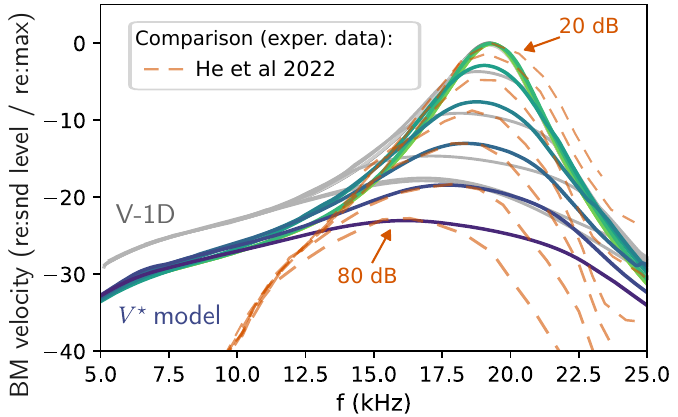}
\caption{BM velocity magnitude normalized to sound level in response to
tones for the \(V^\star\) model (CF=20 kHz). The sound level goes from 20 to 80 dB in 10 dB steps (same color code as in Fig.~\ref{fig:initial_model}). The responses for
the V-1D model (gray lines) and real vibration data (dashed lines) are
reproduced from
Fig.~\ref{fig:initial_model}.}\label{fig:v_star_response}
}
\end{figure}

We simulated the responses to chirps at the CF = 20 kHz place for
the \(V^\star\) model as we did previously for the V-1D model. The
results are shown in Fig.~\ref{fig:v_star_response}. The admittance
correction did not alter the sharpness of the frequency response, but it
extended the compression range by 5 dB. This outcome almost entirely reproduces the desired compression of the peak (seen as the maximum response shifting to lower frequencies with increasing sound levels). The compression of the response
to a CF tone (20 kHz) remains slightly insufficient, falling short by
3 dBs. In addition, the compression achieved by the
model is still associated with a broadening of the cochlear filters, while the
quality factor in the experimental data is only reduced by half
(\(Q_{10dB}=3.4\) at 20 dB vs.~\(Q_{10dB}=1.7\) at 10 dB). As a result,
the frequency responses are substantially broader at high sound levels
compared to those observed experimentally.

\hypertarget{model-calibration-update-rate-and-compressive-growth}{%
\subsection{Model calibration: update rate and compressive
growth}\label{sec:model-calibration}}

Lastly, we present two additional figures related
to the calibration of the \(V^\star\) model.
These figures address two key aspects of the model implementation: the
temporal update of the gain compensation and the calibration of
compressive nonlinearities.

\begin{figure}
\hypertarget{fig:time_response}{%
\centering
\includegraphics{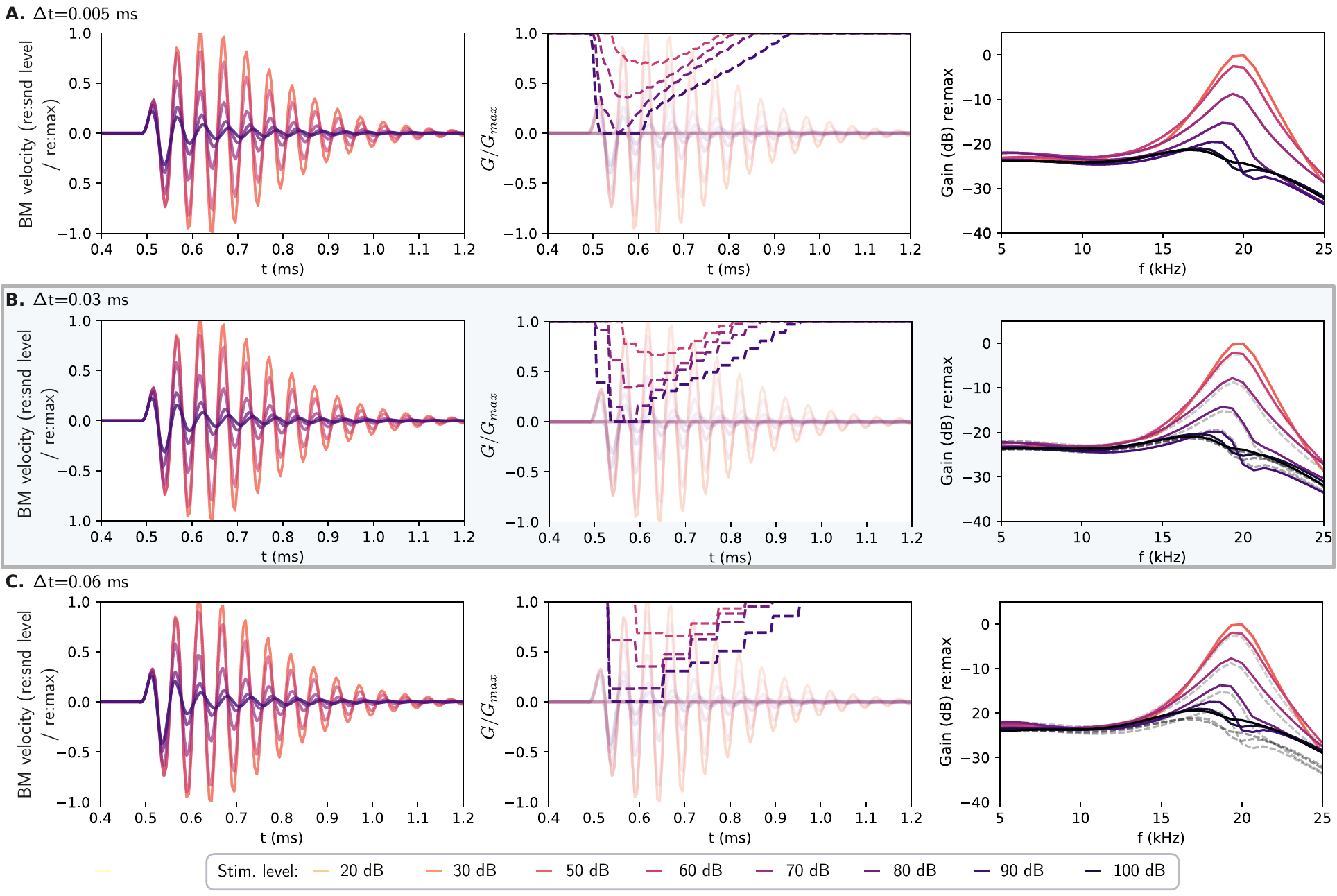}
\caption{\textbf{Responses to clicks for the \(V\star\) model with 
different update rates.} \textbf{Left:} BM velocity responses in the time domain at the CF=20 kHz place, normalized with respect to the stimulus level.
The responses are shown on a scale of {[}-1, 1{]} where 1 corresponds to
the maximal BM response at the most sensitive state. The click was
presented at 0.5 ms. \textbf{Center:} Time course of \(G/G_{max}\)
(dashed curves) representing the strength of the active process used to compute the gain compensation factor. 
In the same figures, the traces of the click response have been
represented (transparent curves). \textbf{Right:} Frequency responses
corresponding to the left panel. The traces in the upper panel are
reproduced in the panels below, in dashed gray, for comparison. Each row
represents a different time resolution for the update of the gain correction: \textbf{A.} 0.005 ms, \textbf{B.} 0.03 ms \textbf{C.}
0.06 ms. The framed row (\(\Delta t = 0.03\) ms) corresponds to the version used in the other figures of the Results section.}\label{fig:time_response}
}
\end{figure}

The factor \(\beta\) in the \(V^\star\) is updated over time based on a
short-term average of output intensity \(\langle I_n \rangle[t]\), which was defined in Methods \ref{sec:update-filter-coefficients}). To illustrate this temporal behavior,
the response to a click at CF=20 kHz is shown in
Fig.\ref{fig:time_response}A. The center panel also shows the time
course of the intermediary variable \(G\) corresponding to the strength of
the active process. \(G\) roughly follows the (vertically flipped) envelope of the BM response, as the active process is at its maximal strength at low intensity levels and inactive at high intensities.
Fig.\ref{fig:time_response}A corresponds to the finer time sampling of
\(G\) allowed by our model (\(\Delta t = \SI{5}{\micro \second}\) or
\(f_s=\SI{200}{\kilo\hertz}\)). This sampling makes the simulations computationally intensive as the \(\beta\) filter
coefficients are re-computed at each time step. With this sampling, the computation times were increased by a factor 20 compared to the V-1D model.
To address this limitation, we tested other versions of the \(V^\star\) model with
slower update rates of the \(\beta\) filters. The corresponding impulse
responses are shown in panels B and C of Fig.\ref{fig:time_response}.
Differences with the faster update (panel A; baseline) are not easily seen on the time
responses, but can be noticed on the frequency responses shown on the
right. With an update rate of 0.06 ms (12 times slower than the fastest
update), the magnitude was different from the baseline by up to \textasciitilde3 dB, which was considered too large. With an update rate of 0.03 ms (6 times slower), deviations
were limited to \textasciitilde1 dB, which was considered as a good trade-off
between computational efficiency and response consistency. The update
rate of 0.03 ms was selected as the default one in our model. It reduces
by almost a factor of 6 the computation time compared to the fastest
update rate, but it still represents around 5 times the computation time
of the V-1D model. However, by storing the filter coefficients in a
lookup table ---computed once at the first execution within a few
seconds---the simulation times become comparable to those of the V-1D
model. These computation times do not include data-saving
operations, which in practice is the main bottleneck of the time-domain
TL model, affecting both V-1D and \(V^\star\) models.

\begin{figure}
\hypertarget{fig:calib_20kHz}{%
\centering
\includegraphics{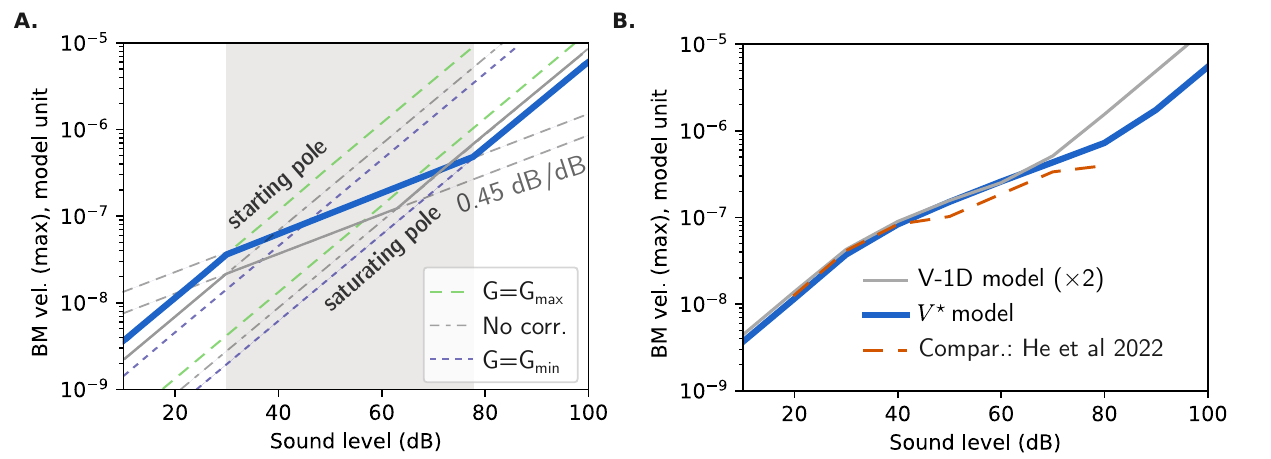}
\caption{\textbf{Calibration of the BM compressive growth}. \textbf{A.}
BM velocity growth functions simulated with different linear versions of
the \(V^\star\) model (dashed lines). The CF of the cochlear section is
20 kHz, and the stimulus probe is a 20 kHz tone. The responses are
divided into two groups: `starting pole' corresponds to a sharply tuned
model, `saturating pole' corresponds to a broadly tuned model adapted to
high sound levels. Green lines were obtained with maximal gain
enhancement (\(G=G_{max}=1.3\)) and dashed blue lines with maximal
gain reduction (\(G=G_{min}=0\)). In addition, we show the
theoretical growth function for the nonlinear \(V^\star\) model as a
solid thick blue line, and for the nonlinear V-1D model (no gain
correction) as a solid gray line. The grayed area highlights the
compressive regime. \textbf{B.} Actual growth functions obtained with
the nonlinear \(V^\star\) and V-1D models. The response for the V-1D
model was vertically shifted (multiplied by a factor 2) to facilitate
comparison. The growth function for the experimental data \citep{He2022}
is also shown (dashed curve).}\label{fig:calib_20kHz}
}
\end{figure}

The calibration of compressive nonlinearities, which was explained at the end of
the Methods section, is illustrated with the gerbil model in Fig.~\ref{fig:calib_20kHz}A. The
procedure is based on the response to a 20 kHz tone at the place with corresponding CF. The response is linear up to a first transition point,
compressive up to a second transition point, and linear again at high
sound levels. In the case of the V-1D model, the compressive regime is
simply the transition between the two linear regimes, corresponding to a
shift from a low-damped to a high-damped system. In the case of the
\(V^{\star}\) model, an additional factor of compression is the pressure
focusing effect, which consists in a gain enhancement at the first
transition point but gradually shifts to a gain reduction over the
compressive regime. Figure~\ref{fig:calib_20kHz} shows the response growth for the two models: panel A illustrates the growth functions expected from the calibration procedure, while panel B shows the actual responses obtained through simulations. The \(V^\star\) model extends
the compressive regime by 10 dB, which is more in line with the
experimental data.

\hypertarget{discussion}{%
\section{Discussion}\label{sec:discussion}}

\hypertarget{improvement-of-compressive-nonlinearities-with-the-vstar-model}{%
\subsection{\texorpdfstring{Improvement of compressive nonlinearities
with the \(V^\star\)
model}{Improvement of compressive nonlinearities with the V\^{}\textbackslash star model}}\label{sec:improvement-of-compressive-nonlinearities-with-the-vstar-model}}

We implemented a method to increase the range of response gain in the Verhulst et
al.~(V-1D) time-domain model, resulting in the modified \(V^\star\)
model. The proposed correction, based on autoregressive filters and
regression methods, introduced an additional source of variation in gain
in the frequency region corresponding to the short-wave region of the
nonlinear traveling wave. The \(V^\star\) model
preserved most characteristics of the V-1D model while providing some
decoupling between frequency selectivity and gain. This was illustrated
in a gerbil model with simulations of BM velocity responses at the CF=20 kHz place. We showed that the compression of the peak response was increased by 5
dB (Fig.\ref{fig:v_star_response}) and the range of sound levels where compression
occurs was extended by 10 dB (Fig.\ref{fig:calib_20kHz}). The limited range of cochlear
response gain was already identified as a potential issue during the
development of the V-1D model, which was originally designed based on
human hearing \citep{Verhulst2018}. The issue appeared more prominent
with recent efforts to adapt the model to small rodents, specifically
gerbils ands mice. The admittance correction proposed in this study
offers a practical solution to compensate for limited compression in
models for species with poor frequency selectivity, but could also be
used to extend the compressive range for species with sharper tuning if
deemed necessary.

However, the proposed method also presented some limitations in achieving a realistic response gain. As seen in the
tip-to-tail ratio in Fig.\ref{fig:v_star_response}, the overall gain of
the simulated filters is well below what is observed experimentally for the BM response. The 5 dB additional gain in the \(V^\star\)
model is not enough to significantly mitigate this issue. While our
method captures the variations in peak gain with changes in sound level,
the baseline gain --- defined as the cochlear response gain in the
passive state --- is inherently limited by the time-domain
implementation. This aspect is further discussed in the next subsection.
This limitation explains why we used the \(\beta\) factor a target for the autoregressive filters, with maximal magnitude between 0.5 and 2
(Fig.\ref{fig:regression_results}), instead of the pressure focusing
factor \(\alpha\) which can contribute approximately 20 dB
of gain at the wave peak \citep{Sisto2021a}. The \(\beta\) factor
captures the variations in the \(\alpha\) factor associated with changes
in strength of the active process, instead of absolute gain. Thus, the
\(V^\star\) model  is mostly relevant when considering a nonlinear model with 
variations of the cochlear response across sound levels. Even when considering only relative differences in gain, we showed that the
regression and filter-design steps tended to reduce the size of the variations of the \(\beta\) factor (Fig.\ref{fig:regression_results}).

Another concern regarding the \(V^\star\) model is the difference in
behavior between the base and apex of the cochlea. The targets for the
\(\beta\) filters were calculated using the S-2D model, which assumes a
constant cochlear height. In reality, the geometry of the cochlea is
tapered, leading to a weaker pressure focusing factor
(Eq.~\ref{eq:p_foc_factor}) at the apex than the base \citep{Altoe2020}.
Therefore, while in practice our method works for CFs below 20 kHz
(e.g., we also analyzed the model response at 4 kHz), its connection to the physics of the cochlea becomes more
questionable at the apex of the cochlea. In the core of the article, we
validated our model against vibration data from the first turn of the
gerbil cochlea \citep{He2022}, where experimental data are relatively
abundant. Existing data measured at the cochlear apex
\citep{Meenderink2022b} suggest that apical responses also exhibit
large compression with limited changes in frequency selectivity. It is
not clear at this point whether the \(V^\star\) model, if adjusted, could faithfully represent the physics of the cochlear apex.

\hypertarget{difficulty-of-achieving-high-gain-in-a-time-domain-tl-model}{%
\subsection{Difficulty of achieving high gain in a time-domain TL
model}\label{sec:difficulty-of-achieving-high-gain-in-a-time-domain-tl-model}}

The previous paragraphs evoked the difficulty of achieving a high tip-to-tail ratio for a time-domain TL model reproducing wide-band BM
responses. To understand why it remains a challenge for the \(V^\star\)
model, it is useful to consider a WKB approximation of a 1-D forward
traveling wave in the form
\(v_{BM}(x, \omega) \propto Y_{BM}^{3/4}(x, \omega) \, \exp(-j \int_0^x \kappa(x', \omega) dx')\)
\citep{Reichenbach2014} \footnote{This formula is valid for a box
model where the fluid impedance \(Z_f = \frac{2 j \omega \rho}{H}\) is
considered constant. For a tapered-box model in the long-wave
approximation, as in Zweig 1991 (hence, as in the V-1D model), the
normalization factor also includes \(1/\sqrt{H}\), which is ignored here as it depends on position but not on \(\omega\).}. Factors contributing to the gain
in \(v_{BM}\) can arise either directly from the admittance \(Y_{BM}\),
or from the exponential term. The limitations of admittance-based gain
have already been mentioned throughout the article: they are tied to the
trade-off between frequency selectivity and gain, well known in the case
of a canonical bandpass filter. Here, we focus instead on how gain could
be achieved through the second factor, the exponential term depending on
the spatial integral of the complex wavelength.

In the sensitive cochlea, a monochromatic wave of pulsation \(\omega\)
is amplified in a small region basal to its peak, where the complex
wavenumber \(\kappa(x', \omega)\) has a positive phase. Assuming the
property of scaling symmetry, this also implies that, at any given
cochlear location \(x\), \(\kappa(x, \cdot)\) exhibits a positive phase
in a narrow frequency range below the BM characteristic frequency
\citep{Altoe2021}. Consequently, \(\kappa(x, \cdot)\), when viewed as a
filter, must include a region where the phase derivative is positive,
corresponding to a negative group delay. This property carries over to
the BM admittance. However, achieving negative group delay for causal
filters is challenging; when it occurs, it results from phase
interference between two wave packets that are necessarily delayed
positively. In practice, a region of negative group delay is possible if
it is compensated by a neighboring frequency region of larger magnitude
with a positive group delay. As an illustration, the \(\beta\) filters
for \(G>1\) in Fig.\ref{fig:regression_results} follow this
pattern, with a negative group delay (positive phase derivative) below
14 kHz followed by a region of positive group delay (negative phase
derivative) above 14 kHz associated with a larger response magnitude.
The limitation described in this paragraph is accentuated by the fact
that a causal filter with a large tip-to-tail ratio is typically
associated with a large positive group delay in the tip region.
Therefore, the two factors of BM velocity gain---either from the BM
admittance or from wave propagation due to a region of negative
resistance---cannot add up in the same frequency band.

It is therefore challenging to design causal filters that combine high tip-to-tail ratios with only modest positive phase shifts in the tip region. This severely limits the time implementation of
a TL model, which relies exclusively on causal components. However, this
raises the question: how does this argument differ for the actual
cochlea? This question is relevant because causality constraints also apply to
physical systems (i.e., Kramers-Koning relations \citep{Shera2007}), yet the \(v_{BM}\) data exhibit a high tip-to-tail ratio. In
reality, causality constraints in the case of the cochlear traveling
wave are not as strict as they may seem initially, because global wave
causality should be considered along with local `point-causality,' with
the former being more flexible than the latter. Local `point-causality'
refers to causal relations between local variables, while global wave
causality also accounts for wave effects that can propagate up to the
speed of the wavefront, appearing quasi-instantaneous to a local
observer. For the time implementation, we have framed the factors
\(\alpha\) or \(\beta\) as part of the updated BM admittance. However, a
more accurate physical interpretation would treat these factors as
distinct from the local BM admittance. Instead, the pressure focusing
factor \(\alpha\) is associated with the forward pressure wave which
separates the frequency components as they travel along the cochlea. The
dispersive nature of the wave means that each wave component is affected
differently by pressure focusing, independently of the delays imposed by
the local BM mechanical response. Consequently, translated as a filter,
\(\alpha\) is not required to be causal, i.e., the translation of local
`point-causality', although it must respect global wave causality. Free
of that constraint, it can significantly and almost instantaneously
enhance the frequency response associated with the short-wave
components, achieving up to +20 dB gain with very short group delays.
This subtle but important distinction highlights the challenge of
adapting a 2-D frequency-domain model into a 1-D time-domain model, a
complexity not immediately apparent from the equations, as it emerges
from a non-trivial interpretation of causality.

The assumption of local causality can even be questioned for the BM
admittance \(Y_{BM}\), at least when it accounts for the BM mechanical
response within the framework of the traveling wave, rather than
treating the BM as an isolated segment. This point is well illustrated
by the S-2D model, in which \(Y_{BM}\) includes a viscous damping term
that depends on the wave profile and is influenced by the dispersive
nature of the wave. More specifically, the viscous stress in the
admittance is related to the second spatial derivative of the pressure
field. When expressed in terms of \(v_{BM}\), it again introduces the
factor \(\alpha\) (Eq.\ref{eq:visc_stress}), proportional to \(k\) in
the short-wave region. A wavelength-dependent admittance is present in
other models: for instance, it is found in the recent work of Elliott et
al.~which includes longitudinal coupling via a viscoelastic model of the
tectorial membrane \citep{Elliott2024}. An admittance model that relaxes
the local point-causality constraint can remove oscillations in the
admittance phase seen in time-domain implementations (at frequencies
below CF) \citep{Shera2001}, allowing for an extended region of negative resistance in the frequency and spatial domains.

Because of the difficulty of simulating a region of negative resistance with a causal and stable model, it
appears necessary for a time-domain implementation to exploit the
resonance of the BM admittance in order to generate a tall response peak. For time-domain models, it is then highly
expected that the wave in response to a CF tone peaks at the cochlear
place where the characteristic frequency is equal to CF, i.e., where the
stiffness and mass terms in the admittance cancel each other. This is
especially true for the V-1D model, as already mentioned in the Methods
section \ref{sec:compar_models}. Although this property appears natural, almost
tautological, it is not necessarily a feature of a 2-D frequency model
(it is not in the S-2D model). If the wavenumber increases sharply in a
region basal to the characteristic place, there can be enough gain
(through \(\alpha\)) and increase in damping (through
viscous damping) so that a tall peak followed by a steep roll-off are
generated in the stiffness-dominated region of the cochlea before the wave reaches the CF place \citep{Sisto2021a, Deloche2024,Sisto2024b}. The occurrence of the traveling-wave peak in the stiffness-dominated region appears more consistent with simultaneous measurements of pressure and BM displacement in the cochlea \citep{Dong2013}, although this question remains a topic of debate.

\hypertarget{two-types-of-nonlinearities-in-the-vstar-model}{%
\subsection{\texorpdfstring{Two types of nonlinearities in the
\(V^\star\)
model}{Two types of nonlinearities in the V\^{}\textbackslash star model}}\label{sec:two-types-of-nonlinearities-in-the-vstar-model}}

The \(V^\star\) model achieves compression through two
different implementation strategies. In the final comments of this
paper, we reflect on these differences and on how this dual approach may prove useful in other contexts.

The solution with the gain enhancement factor \(\beta\) differs from the
initial BM admittance model as it relies on numerical methods rather
than an analytical derivation. The initial V-1D admittance model is
based on a reduced number of parameters (Eq.\ref{eq:shunt_admittance})
which have a direct interpretation in terms of stiffness-to-weight ratio,
damping, magnitude or delay of the slow feedback term. Except for the
slow feedback term, the impedance terms involve time derivatives (up to
second order), corresponding to (quasi-)instantaneous local dynamics. By
contrast, the recursive \(\beta\) filters include a weighted sum of the
previous 32 past values, allowing for slower dynamics. The resulting filter has a smoother, less stereotyped frequency response than the analytical form, but can be adjusted to match a desired broadband shape with more flexibility. We
used as target the pressure focusing factor computed from the S-2D
model, but filter shapes from other models could be used as well. The important point is that the methods presented here are designed to adjust the admittance rather than to reproduce its characteristic form, which in the V-1D model results
from a carefully controlled resonance. The V-1D
admittance equations therefore remain the core of the \(V^\star\) TL model. These equations are constrained by the main poles' trajectory, set to
maintain the near-invariance of BM velocity zero crossings
\citep{Shera2001}. A potential way to generate more compression at CF
would be to lower the local resonant frequency of the BM admittance as
sound level increases. This would accentuate the basal shift of the wave
peak and result in a larger decrease of the response gain to a CF tone with increasing sound levels. However, this approach would also alter
the admittance phase response and disrupt the near-invariance of
zero-crossings. A different way to include pressure focusing in a 1-D model is to integrate the contributions of pressure sources along the cochlear partition using the Green-function formalism \citep{diependaal1989,Duifhuis2012,verhulst2024}, but this approach is computationally very demanding.

Lastly, the two implementations of compressive nonlinearity differ as
they instantiate different types of nonlinear dynamics. Like the V-1D
model, the \(V^\star\) model is primarily driven by instantaneous
nonlinearities, assumed to enhance cochlear motion on a cycle-by-cycle
basis \citep{Dewey2021}. But the gain enhancement factor \(\beta\),
based on a feedback loop, also introduces non-instantaneous
nonlinearities. The strength of pressure focusing follows a vertically-flipped envelope of BM velocity (Fig.~\ref{fig:time_response}). Practically,
this choice speeds up computation since the filter coefficients do not
need to be updated at every timestep. But this choice is also motivated
by a physical argument: a change in the wavenumber cannot be
instantaneous and necessarily requires integration over a time window.
Although most cochlear models include instantaneous nonlinearities,
there has been some debate on whether the active process could instead
follow a feedback loop based on the signal envelope, akin to an automatic
gain control (AGC) scheme \citep{VanDerHeijden2005, Cooper2016, Altoe2017} . In
our model, a part of the nonlinearities are driven by a feedback loop, but these originally arise from a passive phenomenon (pressure
focusing) that accompanies changes in the active process. As a result, the mechanisms underlying the nonlinear compressive response are still considered instantaneous in the case of the \(V^\star\) model. Regardless of its passive or active
origin, the type of nonlinear dynamics in the cochlea has significants 
implications. For example, instantaneous nonlinearities or
envelope-based control loops produce distortion products with different
effect sizes \citep{Duifhuis2012} or phase properties
\citep{VanDerHeijden2005}. In this regard, the \(V^\star\)
model or a variant, with its ability to simulate complex hybrid
dynamics, could prove useful for investigating advanced temporal aspects of
cochlear mechanics.

In conclusion, we presented a method to implement an additional factor of variation of gain response in a 1-D TL model. The factor partly reproduced the response `boost'  observed in the short-wave region in 2-D models. The resulting $V^\star$ model generated 5 dB more gain (or compression) at the wave peak. This method proved useful for developing TL models tailored to the physiology of small mammals, specifically rodents. Developing such models is valuable as part of a translational approach, as common laboratory animals provide extensive insights about normal and impaired hearing. However, achieving this correction required a sophisticated approach involving regression and filter-design methods. Overall, this work highlights the challenge of translating wavelength-dependent phenomena into a time-domain TL implementation while maintaining computational efficiency.

{%
\section*{Acknowledgments}

This project was funded by the FWO project ``Audimod'' G068621N and ERA-NET project ``CoSySpeech'' (G0H6420N).

\appendix

\hypertarget{appendices}{%
\section*{Appendices}\label{sec:appendices}}

\renewcommand{\thesubsection}{\Alph{subsection}. }

\hypertarget{full-form-of-the-cost-function-for-the-lse-regression}{%
\subsection{Full form of the cost function for the LSE
regression}\label{appendix:cost-function}}

The first regression leads to the estimation of coefficients
\([b_k]_{k=1 \cdots 32}\) for a a given pair \((\omega_{BM}, G)\). For
this step, we use a regular discretization of frequencies
\(\left(0 , f_s/(2 m_2) \cdots, m_1 f_s/(2 m_2), \cdots, (m_2-1) f_s \, /(2 m_2) \right)\)
covering the interval \([0, f_s/2]\) where \(f_s/2\) is the Nyquist
frequency for the TL simulations. In our case,
\(f_s=\SI{200}{\kilo \hertz}\) and \(m_2=512\). The set of discrete
frequencies is further divided into
\(\left(0, \cdots, (m_1-1) f_s/(2 m_2) \right)\) and
\(\left(m_1 f_s/(2 m_2), \cdots, (m_2-1) f_s \, /(2 m_2) \right)\) which
correspond to the frequencies below and above
\(\omega_{cut}=1.3\  \omega_{BM}\).

We define the following `feature' matrices:

\begin{subequations}
\begin{align}
X_{l, re} &= \left[\cos\!\left( \pi (k m)/m_2 \right)\right]_{m=0 \cdots m_1-1, \, k=1 \cdots 32},\\
X_{l, im} &= -\left[\sin\!\left( \pi (k m)/m_2 \right)\right]_{m=0 \cdots m_1-1, \, k=1 \cdots 32},\\
X_{r, re} &= \left[\cos\!\left( \pi (k m)/m_2 \right)\right]_{m=m_1 \cdots m_2-1, \, k=1 \cdots 32},\\
X_{r, im} &= -\left[\sin\!\left( \pi (k m)/m_2 \right)\right]_{m=m_1 \cdots m_2-1, \, k=1 \cdots 32} \, .
\end{align}
\end{subequations}

\(l\) stands for `left', \(r\) stands for `right'. Following
Eq.~\ref{eq:approx1}, we have

\begin{subequations}

\label{eq:Y_reg_matrices}
\begin{align}
Y_{l, re}&=\mathrm{Re}(1-1/\beta) \approx 1-\mathrm{Re}(1/\hat\beta) = X_{l, re} b, \\
Y_{l, im}&= \mathrm{Im}(1 - 1/\beta) \approx -\mathrm{Im}(1/\hat{\beta}) = X_{l, im} b, 
\end{align}
\end{subequations}

where \(b\) is the vector of the coefficients \((b_k)\).

The full LSE cost function is then:

\begin{equation}LSE(b)= \left\lVert Y_{l, re} - X_{l, re} b \right\rVert_2^2 + \left\lVert Y_{l, im} - X_{l, im} b \right\rVert_2^2 + \lambda_1 \left\lVert b \right\rVert_2^2 + \lambda_2   \left\lVert \max(0, \frac{X_{r, Im}b}{1-X_{r, Re}b}) \right\rVert_2^2\, \label{eq:full_lse}\end{equation}

where \(\left\lVert \cdot \right\rVert_2\) is the Euclidian norm. The
cost function includes two penalties, weighted by \(\lambda_1\) and
\(\lambda_2\), that prevent large coefficients \(b_k\) and large
positive phase values in \(\hat{\beta}\) for frequencies above
\(\omega_{cut}\). In the second penalty, \(\frac{X_{Im}b}{1-X_{Re}b}\)
corresponds to \(\mathrm{Im}(\hat{\beta})/\mathrm{Re}(\hat{\beta})\),
expressed with the feature matrices and with the division applied
element-wise.

The LSE cost function (Eq.~\ref{eq:full_lse}) was minimized using the
function \texttt{leastsq} of scipy, which is based on the
Levenberg-Marquardt (LM) algorithm.

\subsection{Enforcing null gain of the correction factor for low frequencies}\label{appendix:cost-function-withgain}}

One caveat of the method described above (i.e., using Eq.~\ref{eq:full_lse}) is that the estimated filter  \(\hat\beta(\omega)\) is not guaranteed to be of magnitude 1 when \(\omega \rightarrow 0\). One option to enforce \(\beta(0) = \left(1-\sum_{k=1}^{32} b_k\right)^{-1} = 1\) would be to constrain the coefficients \([b_k]\) to the hyperplane \( \sum_{k=1}^{32} b_k = 0\), but we found that this solution was too rigid, resulting in undesired behaviors. 

Instead, we included a gain factor \(K=1+\epsilon\) in the regression used to determine the filter coefficients, which was adjusted after optimization to enforce a null gain (in dB) at frequency 0.  Specifically, Eq.~\ref{eq:approx1} is replaced with:

\begin{equation}\beta(z^{-1}) \approx \frac{1+\epsilon}{1-\sum_{k=1}^{32} b_k z^{-k}} .\label{eq:approx1withgain}\end{equation}

This updates Eq.~\ref{eq:Y_reg_matrices} to:

\begin{subequations}

\label{eq:Y_reg_matrices_withgain}
\begin{align}
\mathrm{Re}(1-1/\beta) - \epsilon \mathrm{Re}(1/\beta) &\approx X_{l, re} b, \\
\mathrm{Im}(1 - 1/\beta) - \epsilon \mathrm{Im}(1/\beta) &\approx X_{l, im} b, 
\end{align}
\end{subequations}

which leads to the modified LSE function:

\begin{equation}LSE(b, \epsilon)= \left\lVert Y_{l, re} - X_{l, re} b + (Y_{l, re}-1) \epsilon \right\rVert_2^2 + \left\lVert Y_{l, im} - X_{l, im} b +  Y_{l, im} \epsilon \right\rVert_2^2 + \lambda_1 (\left\lVert b \right\rVert_2^2 + \epsilon^2)+ \lambda_2   \left\lVert \max(0, \frac{X_{r, Im}b}{1-X_{r, Re}b}) \right\rVert_2^2\, \label{eq:full_lse_withgain} \ .\end{equation}

The extended LSE cost was minimized as before using a LM algorithm, with \(\epsilon\) as a regressor in addition to the \([b_k]\) coefficients. After the optizimation was done, we set the desired gain of the transfer function \(K=1+\epsilon\) by updating \(\epsilon\) following \(\epsilon \leftarrow -\sum_{k=1}^{32} b_k \).

  \bibliography{biblio.bib}

\begin{thebibliography}{38}
\providecommand{\natexlab}[1]{#1}
\providecommand{\url}[1]{\texttt{#1}}
\expandafter\ifx\csname urlstyle\endcsname\relax
  \providecommand{\doi}[1]{doi: #1}\else
  \providecommand{\doi}{doi: \begingroup \urlstyle{rm}\Url}\fi

\bibitem[Alto{\`e} and Shera(2020)]{Altoe2020}
Alessandro Alto{\`e} and Christopher~A. Shera.
\newblock The cochlear ear horn: Geometric origin of tonotopic variations in
  auditory signal processing.
\newblock \emph{Scientific Reports}, 10:\penalty0 20528, November 2020.
\newblock ISSN 2045-2322.
\newblock \doi{10.1038/s41598-020-77042-w}.

\bibitem[Alto{\`e} et~al.(2014)Alto{\`e}, Pulkki, and Verhulst]{Altoe2014}
Alessandro Alto{\`e}, Ville Pulkki, and Sarah Verhulst.
\newblock Transmission line cochlear models: {{Improved}} accuracy and
  efficiency.
\newblock \emph{The Journal of the Acoustical Society of America}, 136\penalty0
  (4):\penalty0 EL302--EL308, September 2014.
\newblock ISSN 0001-4966.
\newblock \doi{10.1121/1.4896416}.

\bibitem[Alto{\`e} et~al.(2017)Alto{\`e}, Charaziak, and Shera]{Altoe2017}
Alessandro Alto{\`e}, Karolina~K. Charaziak, and Christopher~A. Shera.
\newblock Dynamics of cochlear nonlinearity: {{Automatic}} gain control or
  instantaneous damping?
\newblock \emph{The Journal of the Acoustical Society of America}, 142\penalty0
  (6):\penalty0 3510--3519, December 2017.
\newblock ISSN 0001-4966, 1520-8524.
\newblock \doi{10.1121/1.5014039}.

\bibitem[Alto{\`e} et~al.(2021)Alto{\`e}, Charaziak, Dewey, Moleti, Sisto,
  Oghalai, and Shera]{Altoe2021}
Alessandro Alto{\`e}, Karolina~K. Charaziak, James~B. Dewey, Arturo Moleti,
  Renata Sisto, John~S. Oghalai, and Christopher~A. Shera.
\newblock The {{Elusive Cochlear Filter}}: {{Wave Origin}} of {{Cochlear
  Cross-Frequency Masking}}.
\newblock \emph{Journal of the Association for Research in Otolaryngology
  2021}, pages 1--18, October 2021.
\newblock ISSN 1438-7573.
\newblock \doi{10.1007/S10162-021-00814-2}.

\bibitem[Baby et~al.(2021)Baby, Van Den~Broucke, and Verhulst]{Baby2021}
Deepak Baby, Arthur Van Den~Broucke, and Sarah Verhulst.
\newblock A convolutional neural-network model of human cochlear mechanics and
  filter tuning for real-time applications.
\newblock \emph{Nature machine intelligence}, 3\penalty0 (2):\penalty0
  134--143, February 2021.
\newblock ISSN 2522-5839.
\newblock \doi{10.1038/s42256-020-00286-8}.

\bibitem[Charaziak and Shera(2021)]{Charaziak2021}
Karolina~K. Charaziak and Christopher~A. Shera.
\newblock Reflection-{{Source Emissions Evoked}} with {{Clicks}} and
  {{Frequency Sweeps}}: {{Comparisons Across Levels}}.
\newblock \emph{JARO: Journal of the Association for Research in
  Otolaryngology}, 22\penalty0 (6):\penalty0 641--658, December 2021.
\newblock ISSN 1525-3961.
\newblock \doi{10.1007/s10162-021-00813-3}.

\bibitem[Cooper and {van der Heijden}(2016)]{Cooper2016}
Nigel~P. Cooper and Marcel {van der Heijden}.
\newblock Dynamics of cochlear nonlinearity.
\newblock In Pim {van Dijk}, Deniz Baskent, Etienne Gaudrain, Emile {de
  Kleine}, Andreas Wagner, and Casper Lanting, editors, \emph{Physiology,
  Psychoacoustics and Cognition in Normal and Impaired Hearing}, pages
  267--273. Springer, Berlin, 2016.

\bibitem[Deloche et~al.(2025)Deloche, Thienpont, Moleti, Sisto, and
  Verhulst]{Deloche2024}
François Deloche, Morgan Thienpont, Arturo Moleti, Renata Sisto, and Sarah
  Verhulst.
\newblock Active control of transverse viscoelastic damping in the tectorial
  membrane: {{A}} second mechanism for traveling-wave amplification?
\newblock \emph{Hearing Research}, 464:\penalty0 109320, 2025.
\newblock ISSN 03785955.
\newblock \doi{10.1016/j.heares.2025.109320}.

\bibitem[Dewey et~al.(2021)Dewey, Alto{\`e}, Shera, Applegate, and
  Oghalai]{Dewey2021}
James~B. Dewey, Alessandro Alto{\`e}, Christopher~A. Shera, Brian~E. Applegate,
  and John~S. Oghalai.
\newblock Cochlear outer hair cell electromotility enhances organ of {{Corti}}
  motion on a cycle-by-cycle basis at high frequencies in vivo.
\newblock \emph{Proceedings of the National Academy of Sciences}, 118\penalty0
  (43):\penalty0 e2025206118, October 2021.
\newblock \doi{10.1073/pnas.2025206118}.

\bibitem[Diependaal and Viergever()]{diependaal1989}
Rob~J. Diependaal and Max~A. Viergever.
\newblock Nonlinear and active two‐dimensional cochlear models:
  {{Time}}‐domain solution.
\newblock 85\penalty0 (2):\penalty0 803--812.
\newblock ISSN 0001-4966.
\newblock \doi{10.1121/1.397553}.
\newblock URL \url{https://doi.org/10.1121/1.397553}.

\bibitem[Diependaal et~al.(1987)Diependaal, Duifhuis, Hoogstraten, and
  Viergever]{Diependaal1987}
Rob~J. Diependaal, H.~Duifhuis, H.~W. Hoogstraten, and Max~A. Viergever.
\newblock Numerical methods for solving one-dimensional cochlear models in the
  time domain.
\newblock \emph{The Journal of the Acoustical Society of America}, 82\penalty0
  (5):\penalty0 1655--1666, November 1987.
\newblock ISSN 0001-4966, 1520-8524.
\newblock \doi{10.1121/1.395157}.

\bibitem[Dong and Olson(2013)]{Dong2013}
Wei Dong and Elizabeth~S. Olson.
\newblock Detection of {{Cochlear Amplification}} and {{Its Activation}}.
\newblock \emph{Biophysical Journal}, 105\penalty0 (4):\penalty0 1067--1078,
  August 2013.
\newblock ISSN 0006-3495.
\newblock \doi{10.1016/j.bpj.2013.06.049}.

\bibitem[Duifhuis(2012)]{Duifhuis2012}
Hendrikus Duifhuis.
\newblock \emph{Cochlear {{Mechanics}}: {{Introduction}} to a {{Time Domain
  Analysis}} of the {{Nonlinear Cochlea}}}.
\newblock Springer US, Boston, MA, 2012.
\newblock ISBN 978-1-4419-6116-7 978-1-4419-6117-4.
\newblock \doi{10.1007/978-1-4419-6117-4}.

\bibitem[Elliott et~al.(2024)Elliott, Marrocchio, and Grosh]{Elliott2024}
Stephen Elliott, Riccardo Marrocchio, and Karl Grosh.
\newblock Forms of longitudinal coupling in the organ of {{Corti}}.
\newblock \emph{AIP Conference Proceedings}, 3062\penalty0 (1):\penalty0
  020014, February 2024.
\newblock ISSN 0094-243X.
\newblock \doi{10.1063/5.0189306}.

\bibitem[Fallah et~al.(2021)Fallah, Strimbu, and Olson]{Fallah2021}
Elika Fallah, C.~Elliott Strimbu, and Elizabeth~S. Olson.
\newblock Nonlinearity of intracochlear motion and local cochlear microphonic:
  {{Comparison}} between guinea pig and gerbil.
\newblock \emph{Hearing Research}, 405:\penalty0 108234, June 2021.
\newblock ISSN 03785955.
\newblock \doi{10.1016/j.heares.2021.108234}.

\bibitem[He et~al.(2022)He, Burwood, Porsov, Fridberger, Nuttall, and
  Ren]{He2022}
Wenxuan He, George Burwood, Edward~V. Porsov, Anders Fridberger, Alfred~L.
  Nuttall, and Tianying Ren.
\newblock The reticular lamina and basilar membrane vibrations in the
  transverse direction in the basal turn of the living gerbil cochlea.
\newblock \emph{Scientific Reports}, 12\penalty0 (1):\penalty0 19810, November
  2022.
\newblock ISSN 2045-2322.
\newblock \doi{10.1038/s41598-022-24394-0}.

\bibitem[Meenderink et~al.(2022)Meenderink, Lin, Park, and
  Dong]{Meenderink2022b}
Sebastiaan W.~F. Meenderink, Xiaohui Lin, B.~Hyle Park, and Wei Dong.
\newblock Sound {{Induced Vibrations Deform}} the {{Organ}} of {{Corti
  Complex}} in the {{Low-Frequency Apical Region}} of the {{Gerbil Cochlea}}
  for {{Normal Hearing}} : {{Sound Induced Vibrations Deform}} the {{Organ}} of
  {{Corti Complex}}.
\newblock \emph{Journal of the Association for Research in Otolaryngology:
  JARO}, 23\penalty0 (5):\penalty0 579--591, October 2022.
\newblock ISSN 1438-7573.
\newblock \doi{10.1007/s10162-022-00856-0}.

\bibitem[M{\"u}ller(1996)]{Muller1996}
Marcus M{\"u}ller.
\newblock The cochlear place-frequency map of the adult and developing
  mongolian gerbil.
\newblock \emph{Hearing Research}, 94\penalty0 (1):\penalty0 148--156, May
  1996.
\newblock ISSN 0378-5955.
\newblock \doi{10.1016/0378-5955(95)00230-8}.

\bibitem[Reichenbach and Hudspeth(2014)]{Reichenbach2014}
T.~Reichenbach and A.~J. Hudspeth.
\newblock The physics of hearing: Fluid mechanics and the active process of the
  inner ear.
\newblock \emph{Reports on Progress in Physics}, 77\penalty0 (7):\penalty0
  076601, July 2014.
\newblock ISSN 0034-4885, 1361-6633.
\newblock \doi{10.1088/0034-4885/77/7/076601}.

\bibitem[Ruggero et~al.(1997)Ruggero, Rich, Recio, Narayan, and
  Robles]{Ruggero1998}
Mario~A. Ruggero, Nola~C. Rich, Alberto Recio, S.~Shyamla Narayan, and Luis
  Robles.
\newblock Basilar-membrane responses to tones at the base of the chinchilla
  cochlea.
\newblock \emph{The Journal of the Acoustical Society of America}, 101\penalty0
  (4):\penalty0 2151--2163, April 1997.
\newblock ISSN 0001-4966.
\newblock \doi{10.1121/1.418265}.

\bibitem[Saremi et~al.(2016)Saremi, Beutelmann, Dietz, Ashida, Kretzberg, and
  Verhulst]{Saremi2016}
Amin Saremi, Rainer Beutelmann, Mathias Dietz, Go~Ashida, Jutta Kretzberg, and
  Sarah Verhulst.
\newblock A comparative study of seven human cochlear filter models.
\newblock \emph{The Journal of the Acoustical Society of America}, 140\penalty0
  (3):\penalty0 1618--1634, September 2016.
\newblock ISSN 0001-4966.
\newblock \doi{10.1121/1.4960486}.

\bibitem[Shera(2001)]{Shera2001}
Christopher~A. Shera.
\newblock Intensity-invariance of fine time structure in basilar-membrane click
  responses: {{Implications}} for cochlear mechanics.
\newblock \emph{The Journal of the Acoustical Society of America}, 110\penalty0
  (1):\penalty0 332--348, July 2001.
\newblock ISSN 0001-4966.
\newblock \doi{10.1121/1.1378349}.

\bibitem[Shera(2007)]{Shera2007}
Christopher~A. Shera.
\newblock Laser amplification with a twist: {{Traveling-wave}} propagation and
  gain functions from throughout the cochlea.
\newblock \emph{The Journal of the Acoustical Society of America}, 122\penalty0
  (5):\penalty0 2738--2758, November 2007.
\newblock ISSN 0001-4966, 1520-8524.
\newblock \doi{10.1121/1.2783205}.

\bibitem[Shera and Charaziak(2019)]{Shera2019}
Christopher~A Shera and Karolina~K Charaziak.
\newblock Cochlear frequency tuning and otoacoustic emissions.
\newblock \emph{Cold Spring Harbor Perspectives in Medicine}, 9\penalty0 (2),
  2019.
\newblock ISSN 21571422.
\newblock \doi{10.1101/cshperspect.a033498}.

\bibitem[Shera and Zweig(1991)]{Shera1991}
Christopher~A. Shera and George Zweig.
\newblock A symmetry suppresses the cochlear catastrophe.
\newblock \emph{The Journal of the Acoustical Society of America}, 89\penalty0
  (3):\penalty0 1276--1289, March 1991.
\newblock ISSN 0001-4966, 1520-8524.
\newblock \doi{10.1121/1.400650}.

\bibitem[Shera et~al.()Shera, Guinan, and Oxenham]{Shera2002}
Christopher~A Shera, John~J Guinan, and Andrew~J Oxenham.
\newblock Revised estimates of human cochlear tuning from otoacoustic and
  behavioral measurements.
\newblock 99\penalty0 (5):\penalty0 3318--3323.
\newblock ISSN 0027-8424.
\newblock \doi{10.1073/pnas.032675099}.
\newblock URL \url{http://www.ncbi.nlm.nih.gov/pubmed/11867706}.

\bibitem[Sisto and Moleti(2024)]{Sisto2024b}
Renata Sisto and Arturo Moleti.
\newblock The tonotopic cochlea puzzle: {{A}} resonant transmission line with a
  “non-resonant” response peak.
\newblock \emph{JASA Express Letters}, 4\penalty0 (7):\penalty0 074401, 2024.
\newblock ISSN 2691-1191.
\newblock \doi{10.1121/10.0028020}.

\bibitem[Sisto et~al.(2021)Sisto, Belardinelli, and Moleti]{Sisto2021a}
Renata Sisto, Daniele Belardinelli, and Arturo Moleti.
\newblock Fluid focusing and viscosity allow high gain and stability of the
  cochlear response.
\newblock \emph{The Journal of the Acoustical Society of America}, 150\penalty0
  (6):\penalty0 4283--4296, December 2021.
\newblock ISSN 0001-4966.
\newblock \doi{10.1121/10.0008940}.

\bibitem[Sisto et~al.(2023)Sisto, Belardinelli, Alto{\`e}, Shera, and
  Moleti]{Sisto2023}
Renata Sisto, Daniele Belardinelli, Alessandro Alto{\`e}, Christopher~A. Shera,
  and Arturo Moleti.
\newblock Crucial 3-{{D}} viscous hydrodynamic contributions to the theoretical
  modeling of the cochlear response.
\newblock \emph{The Journal of the Acoustical Society of America}, 153\penalty0
  (1):\penalty0 77--86, January 2023.
\newblock ISSN 0001-4966.
\newblock \doi{10.1121/10.0016809}.

\bibitem[Thienpont et~al.(2024)Thienpont, Deloche, Keshishzadeh, Kiselev,
  Bourien, Puel, Buran, Bramhall, and Verhulst]{Thienpont2024}
Morgan Thienpont, Francois Deloche, Sarineh Keshishzadeh, Danill Kiselev,
  J{\'e}r{\^o}me Bourien, Jean-Luc Puel, Brad Buran, Naomi Bramhall, and Sarah
  Verhulst.
\newblock Translating a {{Computational Model}} of the {{Human Auditory
  Periphery}} to {{Gerbil}} and {{Mouse}} for {{Comparative Auditory
  Research}}.
\newblock In \emph{Mechanics of {{Hearing Workshop}} 2024 ({{MoH}} 24)}, Ann
  Arbor, USA, June 2024. Zenodo.
\newblock \doi{10.5281/zenodo.13769392}.

\bibitem[Van Der~Heijden(2005)]{VanDerHeijden2005}
Marcel Van Der~Heijden.
\newblock Cochlear gain control.
\newblock \emph{The Journal of the Acoustical Society of America}, 117\penalty0
  (3):\penalty0 1223--1233, March 2005.
\newblock ISSN 0001-4966, 1520-8524.
\newblock \doi{10.1121/1.1856375}.

\bibitem[Vecchi et~al.(2022)Vecchi, Varnet, Carney, Dau, Bruce, Verhulst, and
  Majdak]{Vecchi2022}
Alejandro~Osses Vecchi, L{\'e}o Varnet, Laurel~H. Carney, Torsten Dau, Ian~C.
  Bruce, Sarah Verhulst, and Piotr Majdak.
\newblock A comparative study of eight human auditory models of monaural
  processing.
\newblock \emph{Acta Acustica}, 6:\penalty0 17, 2022.
\newblock ISSN 2681-4617.
\newblock \doi{10.1051/aacus/2022008}.

\bibitem[Ver~Hulst()]{verhulst2024}
Henri Ver~Hulst.
\newblock Marrying the physics of critical oscillators with traveling-wave
  models of the cochlea.

\bibitem[Verhulst et~al.(2012)Verhulst, Dau, and Shera]{Verhulst2012}
Sarah Verhulst, Torsten Dau, and Christopher~A. Shera.
\newblock Nonlinear time-domain cochlear model for transient stimulation and
  human otoacoustic emission.
\newblock \emph{The Journal of the Acoustical Society of America}, 132\penalty0
  (6):\penalty0 3842--3848, December 2012.
\newblock ISSN 0001-4966.
\newblock \doi{10.1121/1.4763989}.

\bibitem[Verhulst et~al.(2015)Verhulst, Bharadwaj, Mehraei, Shera, and
  {Shinn-Cunningham}]{verhulst2015}
Sarah Verhulst, Hari~M. Bharadwaj, Golbarg Mehraei, Christopher~A. Shera, and
  Barbara~G. {Shinn-Cunningham}.
\newblock Functional modeling of the human auditory brainstem response to
  broadband stimulation.
\newblock \emph{The Journal of the Acoustical Society of America}, 138\penalty0
  (3):\penalty0 1637--1659, September 2015.
\newblock ISSN 0001-4966.
\newblock \doi{10.1121/1.4928305}.

\bibitem[Verhulst et~al.(2018)Verhulst, Alto{\`e}, and Vasilkov]{Verhulst2018}
Sarah Verhulst, Alessandro Alto{\`e}, and Viacheslav Vasilkov.
\newblock Computational modeling of the human auditory periphery:
  {{Auditory-nerve}} responses, evoked potentials and hearing loss.
\newblock \emph{Hearing Research}, 360:\penalty0 55--75, March 2018.
\newblock ISSN 0378-5955.
\newblock \doi{10.1016/j.heares.2017.12.018}.

\bibitem[Verschooten et~al.(2018)Verschooten, Desloovere, and
  Joris]{Verschooten2018}
Eric Verschooten, Christian Desloovere, and Philip~X. Joris.
\newblock High-resolution frequency tuning but not temporal coding in the human
  cochlea.
\newblock \emph{PLOS Biology}, 16\penalty0 (10):\penalty0 e2005164, October
  2018.
\newblock ISSN 1545-7885.
\newblock \doi{10.1371/journal.pbio.2005164}.

\bibitem[Zweig(1991)]{Zweig1991}
G.~Zweig.
\newblock Finding the impedance of the organ of {{Corti}}.
\newblock \emph{The Journal of the Acoustical Society of America}, 89\penalty0
  (3):\penalty0 1229--1254, March 1991.
\newblock ISSN 0001-4966.
\newblock \doi{10.1121/1.400653}.

\end{thebibliography}

\end{document}